\documentclass[journal,twoside]{IEEEtran}
\usepackage{cite}
\usepackage{bbm, dsfont}
\usepackage{amsmath,amssymb,amsfonts}
\usepackage{algpseudocode}
\usepackage[ruled,linesnumbered]{algorithm2e}
\usepackage{graphicx}
\usepackage{textcomp}
\usepackage{diagbox}
\usepackage{tabularx}
\usepackage{multirow}
\usepackage{capt-of} %
\usepackage{mathtools} %
\usepackage{color, colortbl}
\usepackage{bbding}
\usepackage{pifont}
\usepackage{wasysym}
\usepackage{amssymb}
\usepackage{arydshln}
\usepackage{placeins} 
\usepackage{adjustbox}
\usepackage{nicefrac}
\usepackage{booktabs}
\usepackage{amsmath,amssymb,stmaryrd}   
\usepackage{colortbl}   
\usepackage{xcolor}
\usepackage{soul}

\begin{document}
\title{Weakly-Supervised 3D Medical Image Segmentation using Geometric Prior and Contrastive Similarity }
\author{Hao Du, Qihua Dong, Yan Xu, Jing Liao
\thanks{Hao Du, Qihua Dong and Jing Liao are with Department of Computer Science, 
City University of Hong Kong, Hong Kong, China 
(e-mail: haodu8-c@my.cityu.edu.hk,qihuadong2-c@my.cityu.edu.hk and jingliao@cityu.edu.hk).}
\thanks{Yan Xu are with School of Biological Science and Medical Engineering, Beihang University, Beijing, China 
(e-mail: xuyan04@gmail.com).}
\thanks{Hao Du and Qihua Dong contributed equally to this work. 
Corresponding authors: Jing Liao (jingliao@cityu.edu.hk) and Yan Xu (xuyan04@gmail.com).}
\thanks{This work was supported by the HKSAR Innovation and Technology Commission (ITC) under ITF Project MHP/109/19 and by the National Natural Science Foundation in China under Grant 62022010, the 111 Project
in China under Grant B13003, the high performance computing (HPC)
resources at Beihang University.}
}

\maketitle

\begin{abstract}
Medical image segmentation is almost the most important pre-processing procedure in computer-aided diagnosis but is also a very challenging task due to the complex shapes of segments and various artifacts caused by medical imaging, ({\em i.e.}, low-contrast tissues, and non-homogenous textures). In this paper, we propose a simple yet effective segmentation framework that incorporates the geometric prior and contrastive similarity into the weakly-supervised segmentation framework in a loss-based fashion. The proposed geometric prior built on point cloud provides meticulous geometry to the weakly-supervised segmentation proposal, which serves as better supervision than the inherent property of the bounding-box annotation (\emph{i.e.}, height and width). Furthermore, we propose the contrastive similarity to encourage organ pixels to gather around in the contrastive embedding space, which helps better distinguish low-contrast tissues. The proposed contrastive embedding space can make up for the poor representation of the conventionally-used 
gray space. Extensive experiments are conducted to verify the effectiveness and the robustness of the proposed weakly-supervised segmentation framework. The proposed framework are superior to state-of-the-art weakly-supervised methods on the following publicly accessible datasets: LiTS 2017 Challenge, KiTS 2021 Challenge and LPBA40. We also dissect our method and evaluate the performance of each component.

\end{abstract}

\begin{IEEEkeywords}
Weakly-supervised Segmentation, Medical Image Segmentation, Contrastive Similarity, Geometric Prior, Point Cloud
\end{IEEEkeywords}

\vspace{-0.2cm}
\section{Introduction}
\label{sec:intro}
\IEEEPARstart{S}{egmentation} is of fundamental importance for the understanding and interpretation of medical images, as it is essential for the diagnostic, treatment, and follow-up rehabilitation of various diseases. This task has been widely studied with the recent advent of deep convolutional neural networks (CNNs) ~\cite{ronneberger2015u, isensee2021nnu}. Nevertheless, there exists the main limitation that their methods require a large number of training images with pixel-wise annotations. The extremely high cost of collecting and annotating these training images largely hampers the performance and limits the scalability of deep CNNs in the medical domain. A popular paradigm to alleviate the need for pixel-wise annotations is the weakly-supervised segmentation with bounding-box annotations~\cite{pathak2015constrained,jia2017constrained,kervadec2019constrained,bateson2019constrained, tian2021boxinst}. They employ bounding-box annotations to generate { proposals}, which are {fake} labels and thereby mimic full supervision.

 \begin{figure}[!tb]
    \setlength\tabcolsep{1.6pt}
    \centering
    \small
    \includegraphics[width=\linewidth]{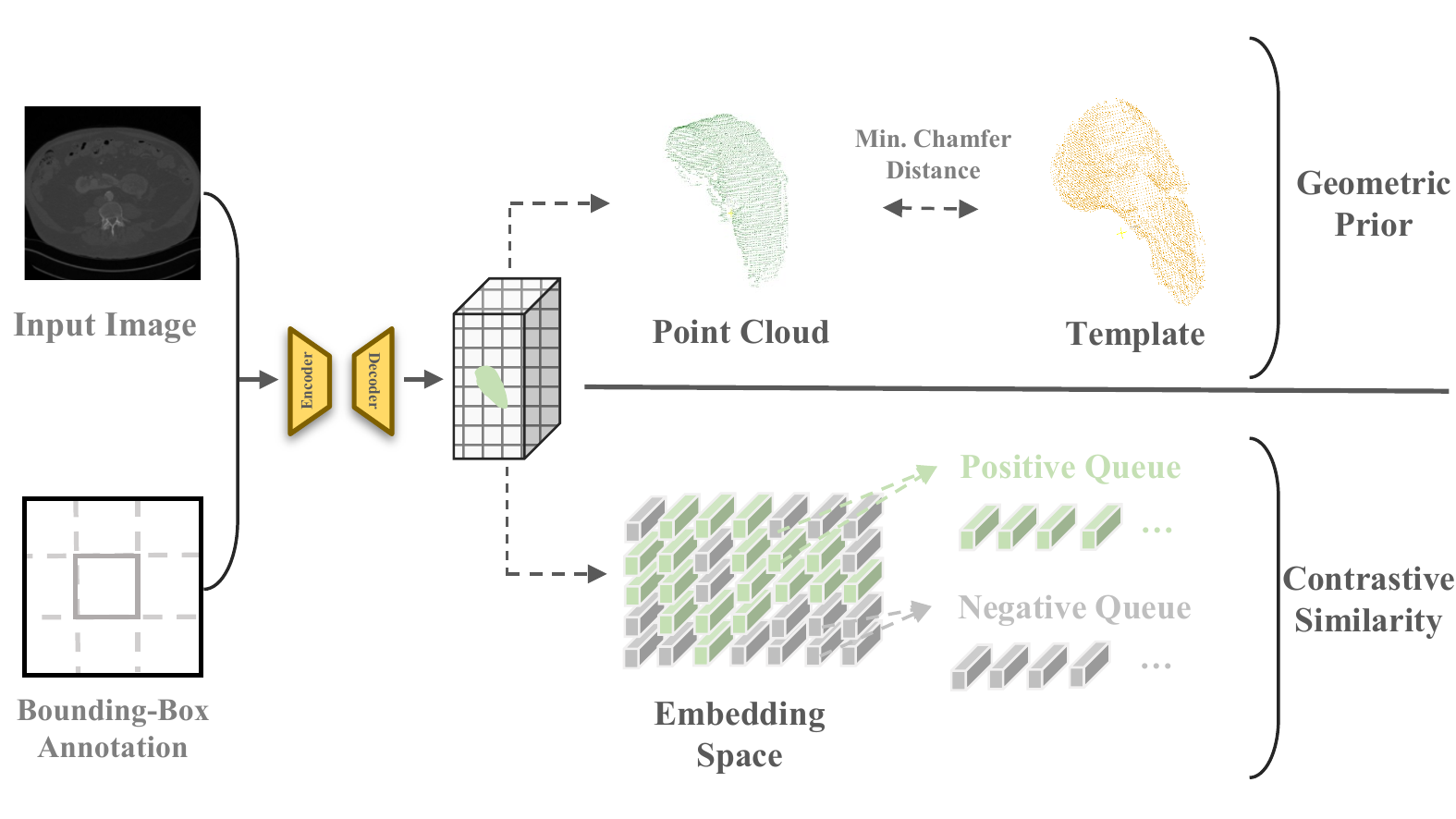}
    \caption{\small{The illustration of the proposed weakly-supervised segmentation framework. As shown in the figure, we propose geometric prior and contrastive similarity for weakly-supervised segmentation. The top row indicates the geometric prior of our method. We first convert the conventionally-used volume representation to point cloud representation and register the template organ to the predicted organ. Then we minimize their Chamfer Distance. The bottom row explains the core idea of the proposed contrastive similarity. By dividing pixels into positive and negative pixels, we encourage organ pixels to gather around in the embedding space to better segment the low-contrast organ.}}
    \label{fig:intro}
    \vspace{-0.3cm}
\end{figure}

\newcommand{\qone}{{complex shapes}}
\newcommand{\qtwo}{{imaging artifacts}}
Nevertheless, despite the good performances achieved by these works in certain practical scenarios, their applicability might be limited for two reasons: 1) {\em \qone}: Some organs have delicate structures {\em, i.e.}, intra-kidney variabilities, which are difficult to be precisely segmented without pixel-wise supervision;
2) {\em \qtwo}: as discussed in previous works~\cite{budrys2018artifacts, boas2012ct,sarkar2018subjective}, various medical imaging artifacts caused by technical or physical problems make low-contrast tissues and non-homogenous textures hard to distinguish, especially in the conventionally widely-used gray space. 
The \qone\ and \qtwo\ largely limit the applicability of the weakly-supervised segmentation models in many scenarios, especially when segmenting complex structures.




To conquer the challenge of \qone\, we propose to learn the geometric prior of the organ by a standard organ template. Instead of using volume representation, we first leverage the gridding reverse~\cite{xie2020grnet} to convert the segmentation result from volume representation to point cloud representation and then compare it with the template in the point cloud space. The basic unit in point cloud representation is much more fine-grained and flexible than the volume representation ({\em, i.e.}, flexible point {\em v.s.} uniform voxel grids), which helps better describe delicate geometric structures. On the other hand, unlike the conventionally-used gray space~\cite{budrys2018artifacts}, we leverage the contrastive learning~\cite{chaitanya2020contrastive} to encode the pixels to high-dimensional embedding space and encourage pixels of the same labels to gather around. This helps alleviate the \qtwo\ for richer expressivity in the embedding space compared to the gray space.



In this paper, we present a novel weakly-supervised segmentation framework, which makes the earliest effort to incorporate geometric prior and contrastive similarity. And the framework is general as well that can be easily applied to improve multiple weakly-supervised segmentation models with bounding-box annotations, \emph{i.e.}, Ai+L~\cite{chu2021improving}, BoxInst~\cite{tian2021boxinst}. By learning geometry prior from the given template and distinguishing low-contrast tissues by the contrastive similarity, our method can generate high-quality results with bounding box supervision only. 

Our method consists of two major components. In the geometric prior component, the shapes of proposals are constrained by a given template represented by point cloud. Both the external boundaries and internal structures of the proposal will be optimized by minimizing the distance according to a given template. The second component is the contrastive similarity, addressing the issues raised by medical imaging artifacts. By pre-training a contrastive head, we successfully learn the difference between organ pixels and non-organ pixels. This component can better distinguish low-contrast tissues and non-homogenous texture than conventionally widely-used gray space. Through extensive experiments, we demonstrate that our method can generate a high-quality segments, along with delicate internal details and accurate boundaries. We show that our method outperforms other bounding-box weakly-supervised methods ~\cite{tian2021boxinst, kervadec2019constrained} under similar settings. We also conduct extensive experiments to verify the effectiveness of components in our method.

In summary, our major contributions are three folds.
\begin{itemize}
    \item We propose a simple yet effective weakly-supervised segmentation framework with bounding-box annotations, which can be easily applied to many weakly-supervised segmentation models and improve their performances.
    \item We propose the geometric prior in point cloud representation to better guide the learning of shapes, especially for those organs with complex structures.
    \item The proposed contrastive similarity makes up for the poor representation of the conventional gray space and thus can better distinguish tissues with medical imaging artifacts.
\end{itemize}
Our code and data will be made publicly available for further research.

\section{Related Work}
\label{sec:related_work}
\IEEEPARstart{I}{n} this section, we first review existing weakly-supervised medical segmentation methods with bounding-box annotations in both natural and medical image segmentation, then we discuss recent works with geometric prior and finally present the trends in contrastive similarity.




\subsection{Weakly-supervised medical semantic segmentation}
Generally, methods in weakly-supervised segmentation are classified into four categories by the type of their weak annotations: scribbles~\cite{lin2016scribblesup}, points~\cite{bearman2016s, Qu2019}, image-level tags~\cite{patel2022weakly, Xu} and bounding-box annotations~\cite{kervadec2019constrained}. Scribbles and points supervision at least label one scribble or point for each region, and the annotated areas will be directly incorporated into the calculation of segmentation loss. Wang \emph{et al.}~\cite{wang2019weakly2} propose to leverage a random walker algorithm~\cite{Grady2006Random} to generate initial proposals for the unlabeled regions and then supervise the training of segmentation models by the initial segments. Qu \emph{{et al.}}~\cite{Qu2019} uses a similar training pipeline but a different label generation method for label generation which combines K-means clustering and Voronoi partition diagram. Xu \emph{et al.}~\cite{Xu} enrich the 
image-level labels to instance-level labels by multiple instance learning (MIL) and segment images using only volume-level labels.



Weakly-supervised segmentation with bounding box annotations earns increasing interest in medical image segmentation for its simplicity and low-annotation cost. We can define the bounding boxes with two corner coordinates that are easy to store in real scenarios. In addition, the bounding box annotations are location-aware so that they provide the spatial relationship of the target object, which is a popular direction in recent researches~\cite{dai2015boxsup,papandreou2015weakly,khoreva2017simple,pu2018graphnet}. In the early stages, researchers~\cite{papandreou2015weakly,rajchl2016deepcut} propose to consider pixels within the bounding box as foreground pixels and train the segmentation framework by these noisy labels. Despite the good performance achieved by such a scheme, they may accumulate errors during the alternative generation process. Most recently, researchers~\cite{tian2021boxinst} tried to directly generate the segmentation result instead of the error-prone alternative way. Generally, they build a {\em mask head} to produce the segmentation result, and the bounding box annotation is employed to train this mask head. 
In this work, we follow this segmentation scheme where the segmentation result is directly generated by the mask head. Furthermore, to address the fore-mentioned \qone\ and \qtwo\, we propose geometric prior and contrastive similarity, respectively.

\vspace{-0.2cm}
\subsection{Geometric prior}
Different from natural images, there exists obvious anatomical prior ({\em, i.e.}, atlas prior) in medical images, specifically in organs of human bodies (\emph{i.e.}, shape and position). Existing works incorporating such anatomical prior mostly fall into two categories: loss-based methods and graph-based methods. Generally, graph-based methods~\cite{patenaude2011bayesian, sabuncu2010generative, fischl2002whole, iglesias2015multi, gao2016segmentation} leverage the probability maps of occurring anatomy chances to construct graph models and estimate the foreground probability from the input image gray space. An appearance model of basic forms is employed to improve the segmentation accuracy~\cite{iglesias2015multi}. Gao {\em et al}~\cite{gao2016segmentation} proposes to apply an appearance ConvNet to characterize the foreground. Despite the high accuracy achieved by these methods, the graphical models bring heavy and expensive computational burdens to the segmentation framework, which makes it infeasible in certain scenarios. 

Another popular direction in combining the prior with the segmentation framework is loss-based methods. Researchers~\cite{ganaye2018semi, bentaieb2016topology, chen2017dcan, zhou2019prior, oktay2017anatomically, dalca2018anatomical} mostly minimize the distance between the segmentation network output and the pre-defined anatomical priors. Several works~\cite{ganaye2018semi, bentaieb2016topology} pose regularization terms on the training objective ({\em, i.e.}, anatomical adjacency or boundary conditions). Distance between predictions and atlas prior are also calculated in latent feature space~\cite{dalca2018anatomical, oktay2017anatomically}.

Compared with graph-based methods, loss-based methods provide a versatile fashion to incorporate anatomical priors with a wider range of scales while maintaining the computational efficiency of the segmentation framework. However, previous loss-based methods fail to address the aforementioned two issues for two aspects: 1) Previous works generally slice the volume into 2D or 3D patches, which are then processed sequentially to save memory cost. However, such a partitioning method breaks the global geometric relationships, resulting in inferior segmentation performance. Different from them, we learn the geometric prior in 3D embedding space to capture the overall geometry and proposed completeness head to ensure the shape completeness of
the proposal. 2) Unlike previous works using a volume representation, we leverage the Gridding Reverse~\cite{xie2020grnet} to convert the volume representation to point cloud representation. Compared to the volume representation constrained by uniform voxel grids, point cloud without grids is more flexible in representing delicate structures.

\vspace{-0.35cm}
\subsection{Contrastive learning}
Contrastive Learning aims to attract the positive and reverse the negative by dividing the feature space into positive and negative data pairs. As for semantic segmentation, it has been mainly used as pre-training~\cite{xie2020pointcontrast,xie2021propagate, he2020momentum}. Van \emph{{et al.}}~\cite{van2021unsupervised} apply it to distinguish features from various salient masks, showing its superiority in unsupervised set-ups. Wang \emph{{et al.}}~\cite{wang2021exploring} have shown advantages of contrastive learning by learning in both pixel
and region levels. Some researchers leverage contrastive learning to address the time-consuming pixel-wise labeling in medical image segmentation. Chaitanya \emph{et al.}~\cite{chaitanya2020contrastive} propose a two-stage
self-supervised contrastive learning framework to learn the feature matching both in global and local mechanisms from unlabeled data in the pre-training stage. Hu \emph{et al.}~\cite{hu2021semi} proposes a semi-supervised scheme to learn self-supervised global contrast and supervised local contrast. In our work, we observe that the conventionally-used gray space is not enough to distinguish positive and negative pixels ({\em, i.e.}, organ pixels and non-organ pixels), especially in Magnetic Resonance Imaging. We thus leverage contrastive learning to calculate the contrastive similarity between pixels. By encoding pixels to high-dimensional features and encouraging pixels of the same label to gather around in the embedding space, it enhances the discriminability and thus alleviates the poor performance of the gray space in handling medical imaging artifacts, \emph{i.e.},  artifacts in ultrasound imaging and similar surrounding tissues. 





\section{Method}
\label{sec:method}

\subsection{Overall framework}
\label{sec:overall}
\newcommand{\R}{\mathbb{R}}

{Given an input image $I \in \R^{S \times H \times W}$ ($S$ indicates the slice number, $H$ indicates height and $W$ represents width) and its corresponding bounding-box annotation $\mathbb{B}^{1\times 6}$ (constrained by its upper left coordinates and bottom right coordinates), our weakly-supervised framework $\mathbf{F}(\cdot)$ obtains the pixel-wise segmentation mask $\mathcal{M}=\mathbf{F}(I)$ and the training goal is to minimize the loss function $L_\text{frame}$:
\begin{equation}
\begin{aligned}
\label{eq:overall_formu}
  \min_{\mathbf{F}}{L_\text{frame}(I, \mathbb{B}, \mathbf{F})}
\end{aligned}
\end{equation}

\textbf{Pipeline} Following nnUNet~\cite{isensee2021nnu}, we randomly sample an input patch $p \in \R^{S' \times H' \times W'}$ from the original input image and encode the patches by a ConvNet encoder $E$. Similar to ~\cite{ronneberger2015u}, we adopt a multi-layer ConvNet as the decoder $G$ to obtain the feature maps $P \in \R^{S' \times H' \times W'}$, where the decoder shares the same layer number as the encoder.} As shown in Fig.~\ref{fig:method}, the proposed geometric prior and contrastive learning are incorporated in the training of the \textit{mask head} to address the aforementioned \qone\ and \qtwo\ issues. The training loss $L_{\text{frame}}$ is composed of two components: $L_{\text{ori}}$ and $L_{\text{mask}}$ 
\begin{equation}
\label{eq:frame_onerall}
\begin{aligned}
    L_{\text{frame}} = L_{\text{ori}} + L_{\text{mask}}
\end{aligned}
\end{equation}
where $L_{\text{ori}}$ indicates the original training loss of a standard weakly-supervised framework ({\em i.e.}, $L_{fcos}$ in BoxInst~\cite{tian2021boxinst}) and $L_{\text{mask}}$ stands for the training loss of the mask head. In the following paragraphs, we mainly discuss the training of the mask head. The training of the mask head can be formulated as Eq.\eqref{eq:overall}. 
\begin{equation}
\label{eq:overall}
\begin{aligned}
    L_{\text{mask}} = L_{\text{geo}} + L_{\text{cons}}
\end{aligned}
\end{equation}
The mask head produce binary segmentation masks which is further optimized by our proposed geometric prior $L_{\text{geo}}$ and contrastive similarity $L_{\text{cons}}$.  More specifically, for geometric prior loss, we build a \textit{completeness head} to predict the completeness score for every proposal, indicating the conditional probability that the object is complete {inside the input}. Each complete proposal is converted into point cloud and registered with the point cloud of the template organ. The Chamfer Distance loss is applied to minimize their distance. In the aspect of the contrastive similarity loss, we build a \textit{contrastive head} to obtain the contrastive similarity by the feature maps that the contrastive head assigns positive and negative labels to each position in the feature maps. 
In the following paragraphs, we first introduce the geometric prior. Then we elaborate on the technical details of the proposed contrastive similarity.


\vspace{-0.35cm}
\subsection{Geometric Prior}



The proposed geometric prior is applied in 3D point cloud space for two reasons: 1) we observe that 2D slices cannot well-preserve the geometric continuity of 3D organs. Thus, we learn the geometric shape of the organ in 3D embedding space. 2) The conventionally-used volume representation cannot handle the segmentation of meticulous structures. The expressivity of the volume representation is largely limited by the uniform voxel grids. Instead, we leverage gridding reverse~\cite{xie2020grnet} to process the segmentation of complex shapes in point cloud embedding space. 

\begin{figure*}[htbp] 
	\centering
	\includegraphics[width=0.9\linewidth]{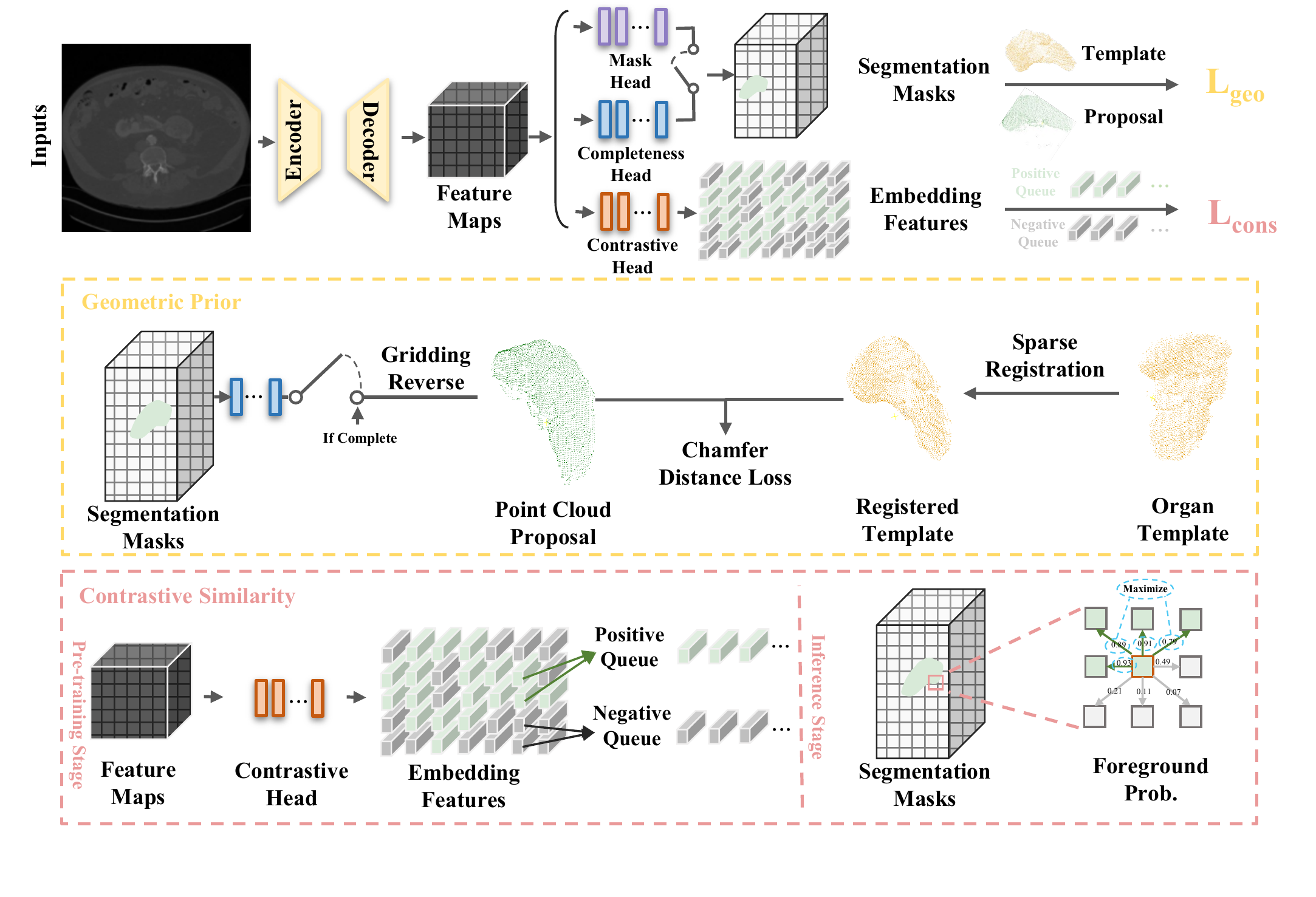}
	\vspace{-0.7cm}
	\caption{The overall framework of our weakly-supervised segmentation framework with geometric prior and contrastive similarity. The training of our framework is organized as follows: We first pre-process the input images and obtain the embedding feature maps. Then we jointly supervise the mask head's training by the proposed geometric prior and contrastive similarity. In the aspect of geometric prior, we first convert the segmentation result to point cloud if complete. Then we calculate the Chamfer Distance between the result and the registered template. In the aspect of contrastive similarity, we first pre-train a contrastive head using only bounding box annotations. Then we minimize the distance between pixels of the same labels by the pre-trained contrastive head.}
	\label{fig:method} 
\end{figure*}

As shown in Fig.~\ref{fig:method}, we propose the geometric prior to better weakly supervise the training of the mask head. The geometric prior refers to the template organ's boundary shape and internal distribution. More specifically, we introduce gridding reverse~\cite{xie2020grnet} to building a bridge between the volume representation and point cloud representation conversion. This helps us to get rid of the representation constraint in the volume representation. After converting both template organ $\mathbb{T}$ and proposal $\mathbb{S}$ into the point cloud representation, we then register the template organ to the proposal by a widely-used ICP registration tool~\cite{Zhou2018}. We finally minimize the geometric prior loss of the Chamfer Distance between the template organ and the proposal in the point cloud embedding space. Below we first introduce the conversion and registration of point cloud and then the definition of geometric prior loss.

\textbf{Conversion \& Registration} To utilize the rich expressivity in
point cloud representation, we introduce the gridding reverse~\cite{xie2020grnet} to help the transition between the volume representation and the point cloud representation. For each voxel grid, the gridding reverse calculates the weighted sum of the eight vertices of the corresponding grid and assigns the weighted sum to coordinates of a new point. Unlike uniform voxel grids, the high flexibility of points' coordinates enable the point cloud representation to describe meticulous and complex architectures. This helps better learn the difficult intra-organ variabilities in the weakly-supervised segmentation. Furthermore, we propose the sparse registration, which is applied before calculating the Chamfer Distance between $\mathbb{S}$ and $\mathbb{T}$. Tiny rotation of the template greatly impacts the calculation of the Chamfer Distance, especially when the object structure is much more complex. We thus conduct registration~\cite{Zhou2018} between the general shape of the proposal and the template. Specifically, we sample 20\% points uniformly across the interval for the template and the proposal respectively, and then calculate the transform matrix between them by the ICP registration tool ~\cite{Zhou2018}. The template point cloud is registered according to the transform matrix.



\textbf{Loss} The geometric prior is then applied in the loss function of the \textit{mask head} training that we optimize the Chamfer Distance between the proposal $\mathbb{S}$ and the registered template organ $\mathbb{T}$. This can be formulated as follows:
\begin{equation}
\begin{aligned}
\label{eq:geo}
    L_{\text{geo}} =  \frac{1}{|\mathbb{S}|}\sum_{x\in \mathbb{S}}\min_{y\in \mathbb{T}}||x-y||_2 + \frac{1}{|\mathbb{T}|}\sum_{y \in \mathbb{T}}\min_{x\in \mathbb{S}}||y-x||_2.
\end{aligned}
\end{equation}
Specifically, the mask head produces binary segmentation masks for each proposal. The proposal is a probabilistic segmentation mask consisting of the segmented instances. This segmentation mask is further processed by the Gumbel-Softmax~\cite{jang2016categorical} to obtain the binary voxel. Locations of low probability are assigned to $0$ and vice versa. We then adopt the gridding reverse~\cite{xie2020grnet} to obtain the point cloud proposal $\mathbb{S}$ from the binary voxel. Finally, we calculate the Chamfer Distance between $\mathbb{S}$ and $\mathbb{T}$ as the geometric prior loss.

\subsection{Contrastive Similarity}
 Gray space performs poorly for the artifacts in medical imaging and similar surrounding tissues. It is not enough to distinguish between positive and negative pixels ({\em, i.e.}, organ pixels and non-organ pixels). We thus leverage contrastive learning to calculate the contrastive similarity between pixels. Encoding pixels to high-dimensional features and encouraging pixels of the same label to gather around in the embedding space helps to increase the distinguishability. 
 




To calculate the proposed contrastive similarity, we first build a ConvNet \textit{contrastive head} after the decoder. Following previous contrastive learning works~\cite{chen2020simple, hu2021semi}, we build a two-layer point-wise convolution $h(\cdot)$ to extract distinct representations from feature maps $P$. More specifically, we first pre-train the proposed \textit{contrastive head} in a coarse-to-fine fashion where only bounding box annotations are included in the pre-training stage. We encode pixels into embedding features $\mathcal{C}=h(P)$ and encourage pixels of the same label to gather around in the embedding space. The contrastive similarity between two pixels is defined as the distance in the embedding space. To calculate the contrastive similarity loss for the whole image, an undirected graph is constructed where the vertices correspond to the pixels and edges are links between neighboring pixels. The contrastive similarity associated with each edge is then summarized for the calculation of the overall contrastive similarity loss. Below we first introduce the pre-training of the contrastive head and then the definition of contrastive similarity loss.


\textbf{Pre-training} The pre-training of the contrastive head is conducted in a weakly-supervised fashion that only bounding box annotations are included in the pre-training stage. To be more specific, there are two sub-stages in the pre-training stage: {\em coarse} and {\em refine}. In the {\em coarse} stage,
we first take pixels within the bounding box as positive labels and pixels outside the bounding box as negative labels. Then we train the contrastive head by such labeling. 
However, the performance of the contrastive head is largely limited for the noisy labels. Thus, we propose the {\em refine} stage to further improve the performance of the contrastive head.
In the {\em refine} stage, we first take random $K$ negative pixels as {\em referring pixels}. And for each pixel within the bounding box, we calculate the distance $\mathcal{D}$ between all K referring pixels:
\begin{equation}
\begin{aligned}
\label{eq:cons1}
    D_{u, v, z} = \sum_{i=1}^{K}{\mathbbm{1}\{\{\mathcal{C}_{u,v,z} \cdot \mathcal{C}_{i}\}\geq \tau\} }
\end{aligned}
\end{equation}
where $\mathcal{C}_{u,v,z}$ $\in$ $R^{S \times H \times W}$ indicates the feature at the $(u, v, z)$ of the embedding features, $\mathbbm{1}$ stands for 1 if the distance is greater than $\tau$ and 0 if less, and $\tau$ is the threshold to decide whether pixels are positive or negative. If $D_{u, v, z}$ is greater than $K/2$, the pixel at location $(u, v, z)$ is considered positive, and vice versa. Then, we train the contrastive head using the same training loss as in the {\em coarse} stage, which is formulated as:
\begin{equation}
\begin{aligned}
\label{eq:cons2}
    loss= -\frac{1}{\left | \Omega \right |}\sum_{(u, v, z) \in \Omega}\frac{1}{\left | \mathcal{P}(u, v, z) \right |} \cdot
    ~~~~~~~~~~~~~~~~~~&\\
    ~~~~log\frac{\sum_{(u_p, v_p, z_p) \in \mathcal{P}(u, v, z)}exp(\mathcal{C}_{u ,v, z}\cdot \mathcal{C}_{u_p, v_p, z_p}/\tau )}{\sum_{({u}_{n}, {v}_{n}, {z}_{n}) \in \mathcal{N}(u, v, z)}exp(\mathcal{C}_{u, v, z}\cdot \mathcal{C}_{{u}_{n}, {p}_{n}, z_n}/\tau)}
\end{aligned}
\end{equation}
where $\mathcal{C}_{u,v,z}$ $\in$ $R^{S \times H \times W}$ indicates the feature at the $(u, v, z)$ of the feature map, and
$\Omega$ stands for {all points inside input}. 
$\mathcal{P}(u, v, z)$ denotes the set of points with the same label as the pixel at $(u, v, z)$  and $\mathcal{N}$ denotes the set of points with different labels. $\tau$ is the temperature constant.

\textbf{Loss} Considering an undirected graph $G=(V, E)$ built on the input image $I$, where $V$ corresponds pixels and $E$ indicates edges between neighboring pixels, the predicted segmentation mask can be viewed as the probability of pixel $(u, v, z)$ being foreground. Then the probability of pixel $(u_1, v_1, z_1)$ and pixel $(u_2, v_2, z_2)$ being the same label is: 
\begin{equation}
\label{eq:prob}
\begin{aligned}
Prob(y_e = 1) = \tilde{\mathcal{M}}_{u_1, v_1, z_1}
     \cdot
     {\tilde{\mathcal{M}}}_{u_2, v_2, z_2} 
     ~~~~~~~~~~~~&~~~~~~~~~~~~\\~~~~~~~~~~~~~~~+ (1 - {\tilde{\mathcal{M}}}_{u_1, v_1, z_1})     \cdot(1 - \tilde{\mathcal{M}}_{u_2, v_2, z_2}),
\end{aligned}
\end{equation}
where $\tilde{\mathcal{M}}$ indicates the foreground probability mask and $y_e$ represents the label of the edge.

Thus, we can define an indicator on each edge to indicate whether they belongs to the same label. If the contrastive similarity between two neighboring pixels is above the pre-defined threshold $\tau$, the indicator on the edge linking them is assigned to $1$, and $0$ vice versa. We discard the edges with $0$ and further summarize the contrastive similarity loss of positive edges, which can be formulated as:
\begin{equation}
\label{eq:sim_loss}
\begin{aligned}
L_{\text{cons}} = -\frac{1}{N}\sum_{e \in E_{in}}\mathbbm{1}_{\{\mathcal{C}_{e_{start}} \cdot \mathcal{C}_{e_{end}} \geq \tau\}}\log Prob(y_e = 1).
\end{aligned}
\end{equation}
This serves as the contrastive similarity loss of the whole image.

\begin{table*}[t!]
\small
\centering
\vspace{-0.3cm}
\caption{Performance comparison with other methods. We evaluate the performance using two metrics: Dice Score and HD95. $\dagger$ we obtain their dice score by their open source code. $\ddagger$: we rebuild the 3D framework based on its 2D open source code. {\em Ours w/o geometric} indicates we remove the geometric prior component. {\em Ours w/o contrastive} indicates we remove the contrastive similarity component.}
\renewcommand\arraystretch{1.4}
\newcolumntype{P}[1]{>{\centering\arraybackslash}p{#1}}
\resizebox{1.\linewidth}{!}{
    \begin{tabular}{P{4.3cm}P{2.7cm}P{1.3cm}P{1.3cm}cP{1.3cm}P{1.3cm}cP{1.3cm}P{1.3cm}}
        \specialrule{.8pt}{0pt}{2pt}
        \multirow{2}{*}{Method}  &  \multirow{2}{*}{Backbone} &
        \multicolumn{2}{c}{LiTS17}  && \multicolumn{2}{c}{KITS21}  && \multicolumn{2}{c}{LPBA40}  \\ 
        \cline{3-4}  \cline{6-7} \cline{9-10}
           &   &  $\uparrow$DSC(\%)&$\downarrow$HD95&&$\uparrow$DSC(\%)&$\downarrow$HD95  && $\uparrow$DSC(\%)  & $\downarrow$HD95  \\ 
          \specialrule{.4pt}{2pt}{0pt}
          Fully Supervised & UNet~\cite{ronneberger2015u} & 95.5 & 5.3 && 96.0 & 3.2 && 83.7 & 2.1 \\
          \hdashline
          DeepCut (2016)~\cite{rajchl2016deepcut} & - & 37.1 & 15.2 && 36.2 & 14.7 && - & - \\
          SDI (2017)~\cite{khoreva2017simple} & VGG-16~\cite{simonyan2014very} & 49.2 & 11.7 && - & - && 38.7 & 9.0 \\
          
           MIL (2020)~\cite{Dolz2020}$^{\dagger}$  & ENet~\cite{paszke2016enet}& 69.4 & 9.4 && 72.3 & 8.7 &&- & - \\
           GMIL (2021)~\cite{hu2021semi}$^{\dagger}$  & ENet~\cite{paszke2016enet} & 71.1 & 8.8 && 71.7 & 5.9 && 58.2& 4.7\\
          
          BoxInst (2021)~\cite{tian2021boxinst}$^{\ddagger}$  &   UNet~\cite{ronneberger2015u} & 47.1 & 11.3 && 48.4 & 11.6 && 37.9 & 8.9\\
          \rowcolor[gray]{.95}
          Ours {\text{w/o geometric}} & UNet~\cite{ronneberger2015u} & {52.9}& {10.9} && {54.2} &{10.7}  && {44.9} & {7.1}\\
          \rowcolor[gray]{.95}
          Ours {\text{w/o contrastive}} & UNet~\cite{ronneberger2015u} & {69.7}&{9.4} && {68.5} &{9.1} && {57.3} & {4.9}\\
          \rowcolor[gray]{.9}
          Ours & UNet~\cite{ronneberger2015u} & \textbf{79.8} &\textbf{8.7} && \textbf{80.2}& \textbf{5.3} && \textbf{65.4} & \textbf{4.2}\\
        
        \specialrule{.8pt}{0pt}{2pt}
    \end{tabular}
}
\label{tab:main_res}
\vspace{-0.6cm}
\end{table*}




\section{Experiments}
\label{sec:exp}

\subsection{Datasets}
\begin{itemize}
    \item LiTS: The public liver LiTS~\cite{bilic2019liver} dataset comprises 201 CT scans from various CT scanners and devices. The resolution of images in this dataset is from $0.56$mm to $1.0$mm in axial and 0.45mm to 6.0mm in $z$ direction. Slices in $z$ range from 42 to 1026. We split 131 cases into training and evaluation sets by a ratio of 4:1.
    \item KiTS21: The publicly accessible KiTS21~\cite{heller2020state} dataset consists of 300 cases during the period from 2010 to 2020. Each CT scan in the dataset is annotated by three expert annotators for the following semantic classes: Kidney, Tumor, and Cyst. We split the provided CT scans into training and validation sets by a ratio of 4:1. 
    \item LPBA40: This dataset consists of 40 T1-weighted image volumes from randomly selected cases (among 40 scans: 20 males, 20 females, and 29.2 $\pm$ 6.3 years). 
    The scans were acquired with a spatial resolution of 0.86 $\times$ 1.5 $\times$ 0.86 mm$^3$. Here, we conduct experiments on the subset of the hippocampus in LPBA40.
.
\end{itemize}

\subsection{Implementation details and evaluation metrics}
The proposed framework with geometric prior and contrastive similarity can be easily incorporated with any weakly-supervised segmentation models with bounding box annotations. To verify the robustness of our method, we conduct experiments with two models: Ai+L~\cite{chu2021improving} and BoxInst~\cite{tian2021boxinst}. Here, we take the latest BoxInst~\cite{tian2021boxinst} as our baseline, and experiments are conducted based on this model unless otherwise specified.
The training setting is mostly based on BoxInst's training settings: The basic learning rate is 0.01 with weight decay 1e-4, and a MultiStepLR scheduler with warmup is adopted. Our framework is trained on GeForce RTX 3090 GPU. During the inference stage, we set the threshold of the completeness head to 0.6 and the threshold of the class head to 0.5 empirically. We randomly select a training sample as the template for each specific dataset unless otherwise specified. We evaluate our results by two widely adopted metrics: the Dice score (DSC) in percentages and the Hausdorff Distance (95\%). 1) DSC score is calculated as the overlap area of two masks divided by their summation. A higher DSC score corresponds to better overlap with GT. 2) The Hausdorff Distance mainly measures the boundary distance between the segmentation result and the pixel-wise segmentation masks. Better segmentation results are of a smaller value than inferior results.

\textbf{Pre-processing \& Post-processing} Following nnUNet~\cite{isensee2021nnu}, the pre-processing includes downsampling, patching, and data augmentation. Here, we downsample the data to reduce memory use and ensure the existence of complete instances inside one patch. Then in the post-processing stage, the pipeline contains resampling, patching, and patch-NMS (Non-Maximum Suppression between patches). We first adjust the spacing and patch the data to the same size as training, with a certain step size. Then we predict a segmentation mask for each patch and return all the predictions to the original space. Since there are overlapping areas between patches, we use patch-NMS in these areas to eliminate the duplicates. 

\textbf{Bounding box annotation} We conduct experiments on LiTS, KiTS, and LPBA40 datasets. All these three datasets have pixel-wise segmentation annotations. We {utilize} the corresponding pixel-wise annotations to obtain the corresponding bounding-box annotations. In the training stage, only bounding-box annotations are included. During the inference stage, we evaluate the performance by pixel-wise annotations.

\subsection{Results}

\textbf{Quantitative Results.} We report the quantitative results of three datasets (LiTS17, KiTS21, and LPBA40) in Tab.~\ref{tab:main_res}. 
Our method aims to integrate the geometric prior and contrastive similarity to give better supervision in the training of the weakly-supervised segmentation framework. Compared with BoxInst~\cite{tian2021boxinst}, which is the baseline of our method, the segmentation performance of our method is largely improved over all three datasets. This is because the geometric prior can supervise the learning of both outer shape and the inner structure, and the contrastive similarity better distinguishes organ and non-organ pixels in the embedding space. Similarly, ours outperforms the most recent two weakly-supervised methods~\cite{hu2021semi, Dolz2020} supervised by bounding-box annotations, which are based on the Multiple Instance Learning (\emph{i.e.}, MIL). The reason behind this is that MIL mainly focuses on the boundary regression of the proposal, but additionally, we learn the internal details of organs from the geometric prior. We also evaluate the upper bound of our method in which we supervise the framework training with ground truth. 

\begin{figure*}[ht]
    \setlength\tabcolsep{1.1pt}
    \centering
    \begin{tabularx}{\textwidth}{ccccccccc}
    &&\text{Input} &\text{GT}& \text{Fully} & \text{BoxInst} & \text{MIL} & \text{GMIL} & \text{Ours} \\
    
    \raisebox{0.65\height}{\rotatebox{90}{\text{Kidney}}}&
    \raisebox{0.32\height}{\rotatebox{90}{\text{Surface X}}}&
    \includegraphics[width=0.13\linewidth]{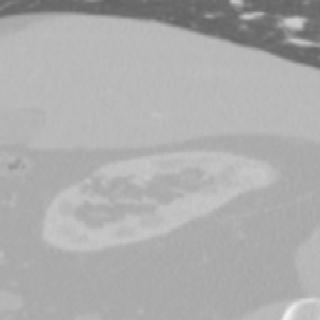}&
    \includegraphics[width=0.13\linewidth]{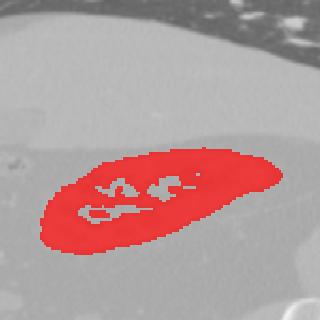}&
    \includegraphics[width=0.13\linewidth]{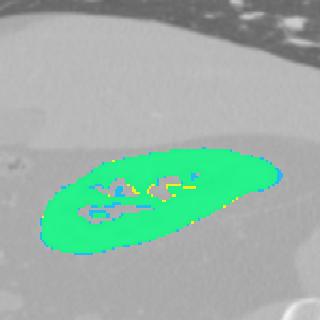}&
    \includegraphics[width=0.13\linewidth]{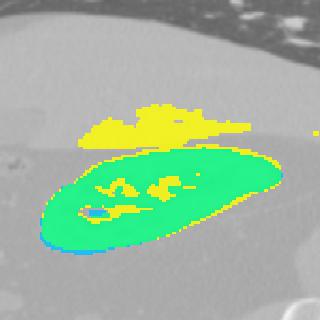}&
    \includegraphics[width=0.13\linewidth]{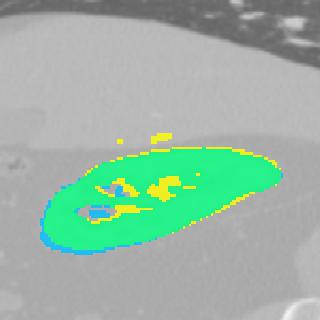}&
    \includegraphics[width=0.13\linewidth]{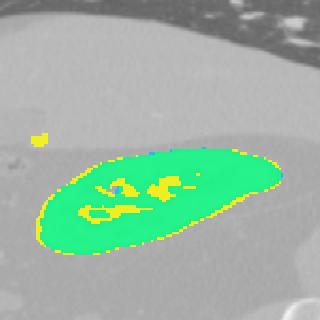}&
    \includegraphics[width=0.13\linewidth]{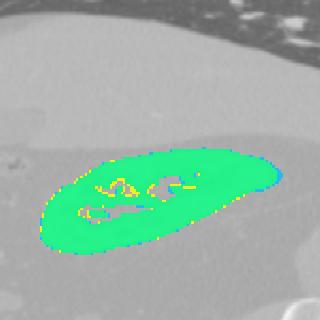}\\
    
    \raisebox{0.65\height}{\rotatebox{90}{\text{Kidney}}}&
    \raisebox{0.32\height}{\rotatebox{90}{\text{Surface Y}}}&
    \includegraphics[width=0.13\linewidth]{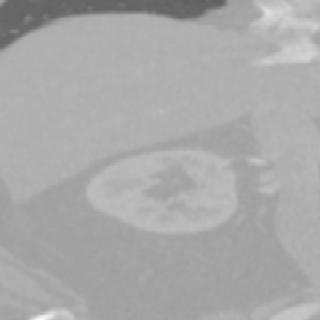}&
    \includegraphics[width=0.13\linewidth]{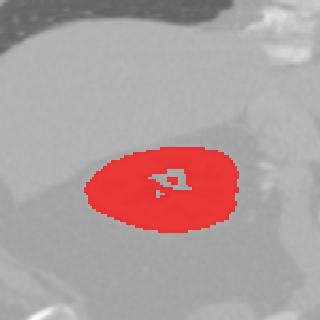}&
    \includegraphics[width=0.13\linewidth]{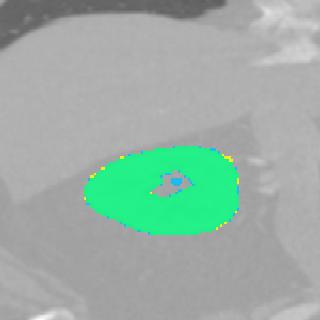}&
    \includegraphics[width=0.13\linewidth]{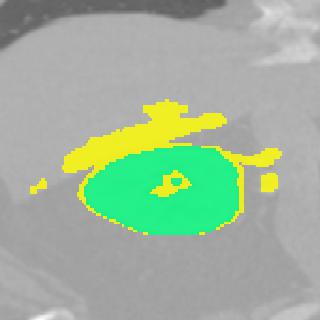}&
    \includegraphics[width=0.13\linewidth]{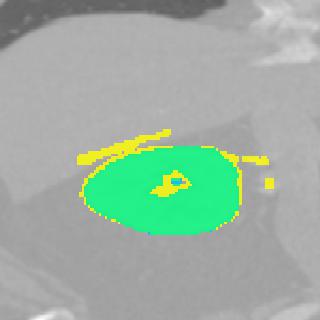}&
    \includegraphics[width=0.13\linewidth]{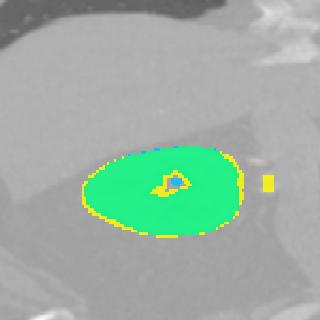}&
    \includegraphics[width=0.13\linewidth]{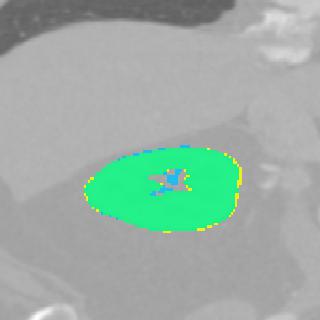}\\
    
    \raisebox{0.65\height}{\rotatebox{90}{\text{Kidney}}}&
    \raisebox{0.32\height}{\rotatebox{90}{\text{Surface Z}}}&
    \includegraphics[width=0.13\linewidth]{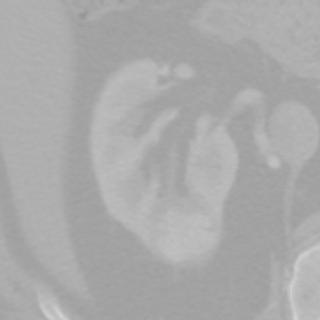}&
    \includegraphics[width=0.13\linewidth]{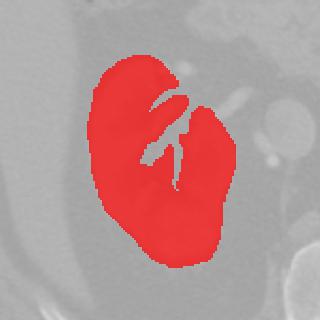}&
    \includegraphics[width=0.13\linewidth]{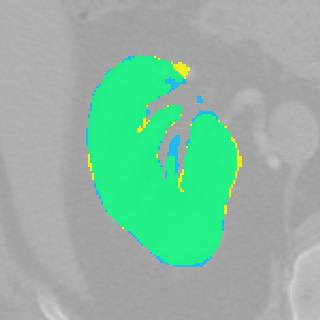}&
    \includegraphics[width=0.13\linewidth]{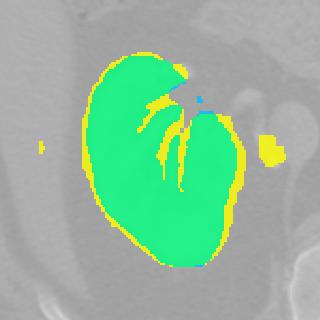}&
    \includegraphics[width=0.13\linewidth]{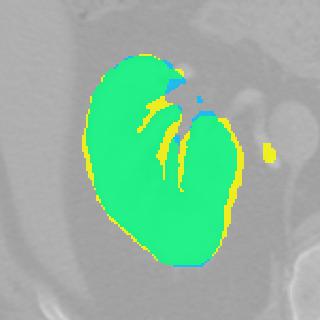}&
    \includegraphics[width=0.13\linewidth]{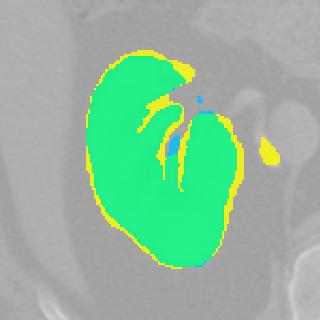}&
    \includegraphics[width=0.13\linewidth]{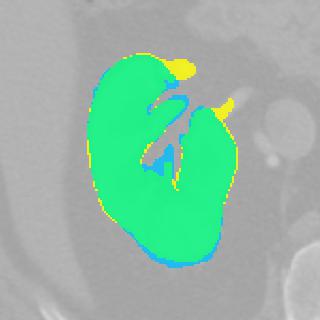}\\
    
    \raisebox{1.1\height}{\rotatebox{90}{\text{Liver}}}&
    \raisebox{0.32\height}{\rotatebox{90}{\text{Surface X}}}&
    \includegraphics[width=0.13\linewidth]{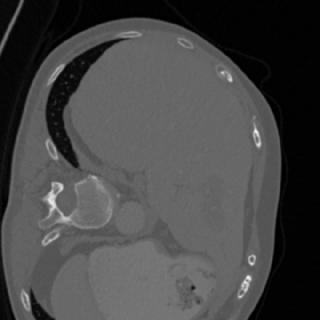}&
    \includegraphics[width=0.13\linewidth]{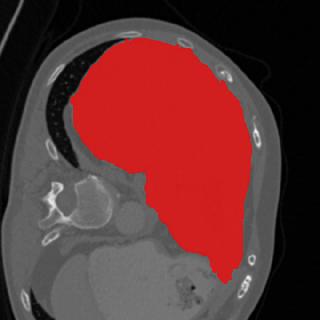}&
    \includegraphics[width=0.13\linewidth]{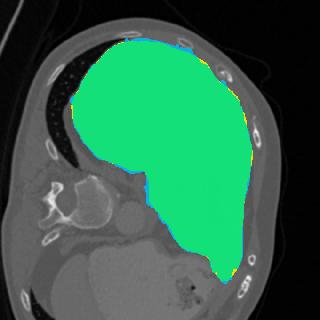}&
    \includegraphics[width=0.13\linewidth]{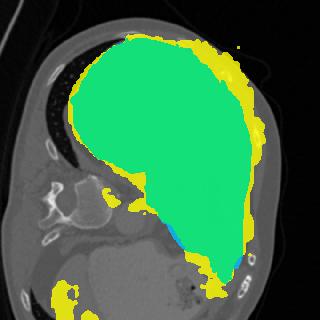}&
    \includegraphics[width=0.13\linewidth]{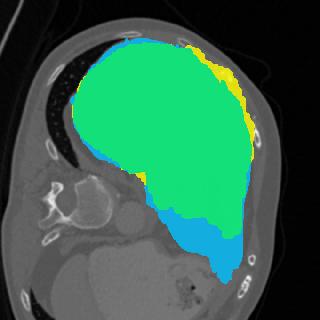}&
    \includegraphics[width=0.13\linewidth]{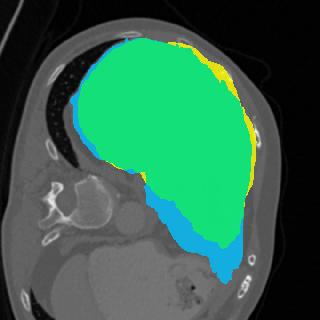}&
    \includegraphics[width=0.13\linewidth]{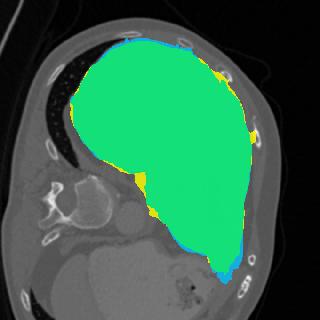}\\
    
    \raisebox{1.1\height}{\rotatebox{90}{\text{Liver}}}&
    \raisebox{0.32\height}{\rotatebox{90}{\text{Surface Y}}}&
    \includegraphics[width=0.13\linewidth]{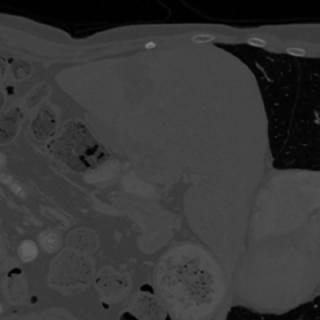}&
    \includegraphics[width=0.13\linewidth]{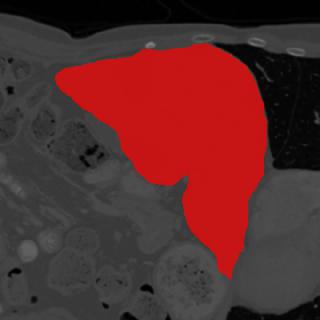}&
    \includegraphics[width=0.13\linewidth]{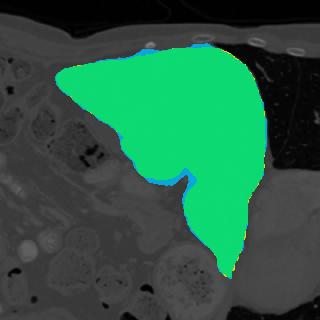}&
    \includegraphics[width=0.13\linewidth]{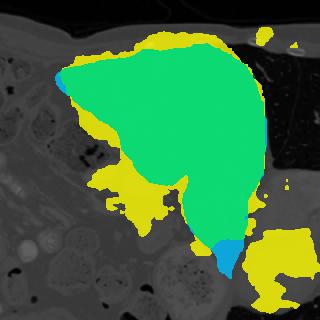}&
    \includegraphics[width=0.13\linewidth]{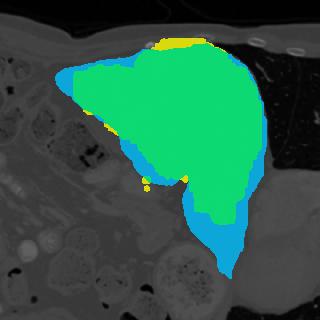}&
    \includegraphics[width=0.13\linewidth]{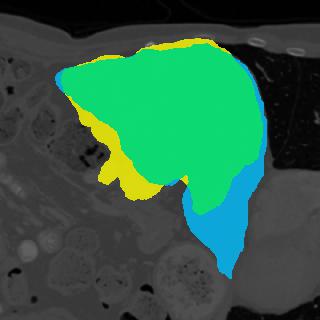}&
    \includegraphics[width=0.13\linewidth]{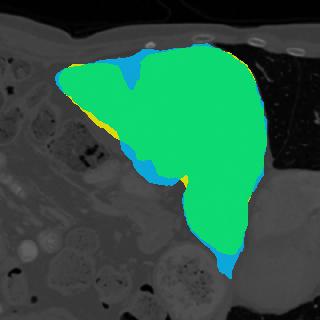}\\
    
    \raisebox{1.1\height}{\rotatebox{90}{\text{Liver}}}&
    \raisebox{0.32\height}{\rotatebox{90}{\text{Surface Z}}}&
    \includegraphics[width=0.13\linewidth]{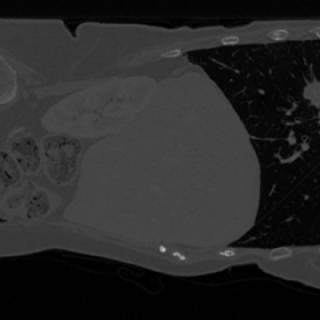}&
    \includegraphics[width=0.13\linewidth]{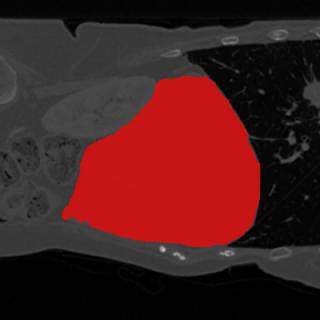}&
    \includegraphics[width=0.13\linewidth]{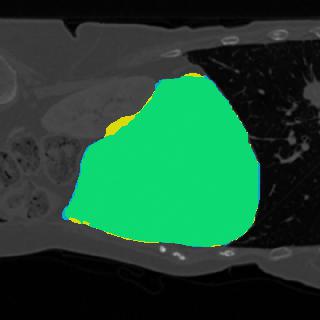}&
    \includegraphics[width=0.13\linewidth]{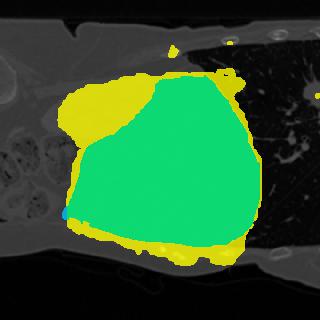}&
    \includegraphics[width=0.13\linewidth]{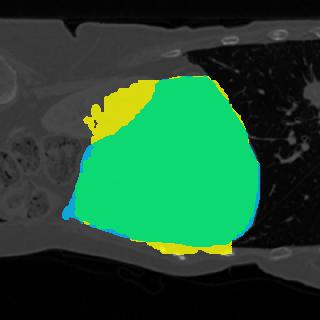}&
    \includegraphics[width=0.13\linewidth]{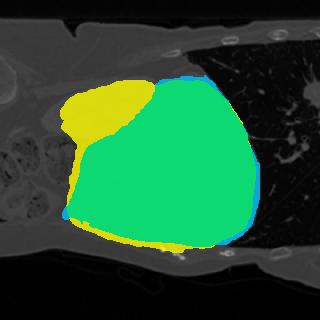}&
    \includegraphics[width=0.13\linewidth]{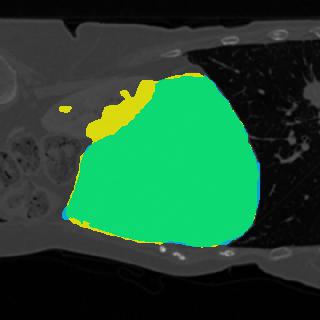}\\
    
    \raisebox{0.1\height}{\rotatebox{90}{\text{Hippocampus}}}&
    \raisebox{0.32\height}{\rotatebox{90}{\text{Surface X}}}&
    \includegraphics[width=0.13\linewidth]{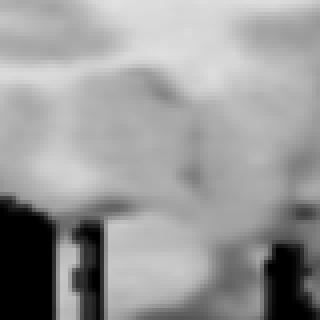}&
    \includegraphics[width=0.13\linewidth]{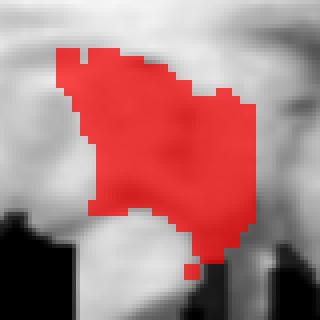}&
    \includegraphics[width=0.13\linewidth]{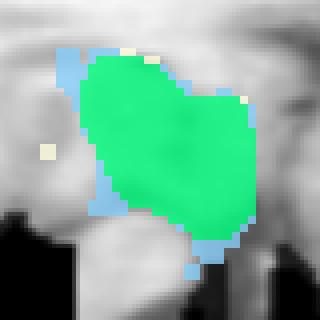}&
    \includegraphics[width=0.13\linewidth]{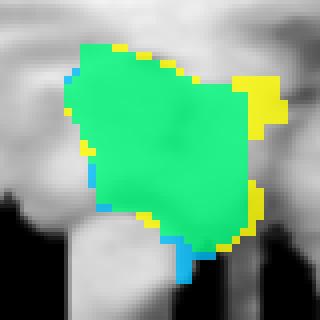}&
    \includegraphics[width=0.13\linewidth]{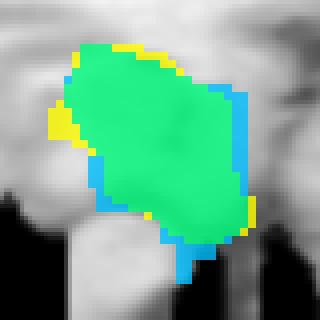}&
    \includegraphics[width=0.13\linewidth]{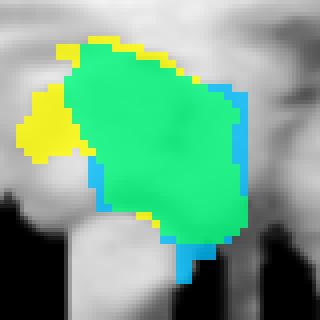}&
    \includegraphics[width=0.13\linewidth]{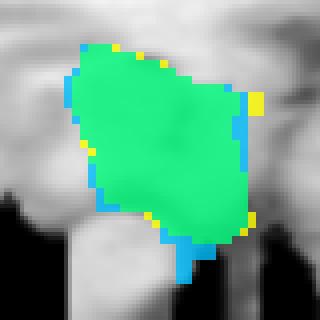}\\
    
    \raisebox{0.1\height}{\rotatebox{90}{\text{Hippocampus}}}&
    \raisebox{0.32\height}{\rotatebox{90}{\text{Surface Y}}}&
    \includegraphics[width=0.13\linewidth]{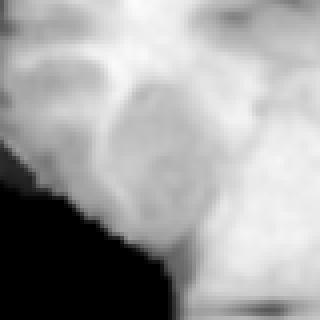}&
    \includegraphics[width=0.13\linewidth]{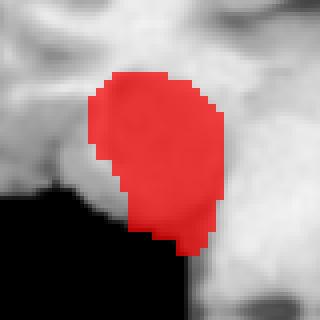}&
    \includegraphics[width=0.13\linewidth]{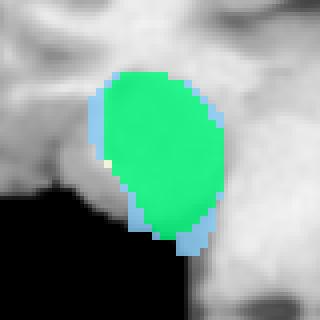}&
    \includegraphics[width=0.13\linewidth]{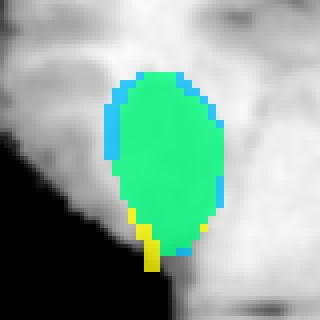}&
    \includegraphics[width=0.13\linewidth]{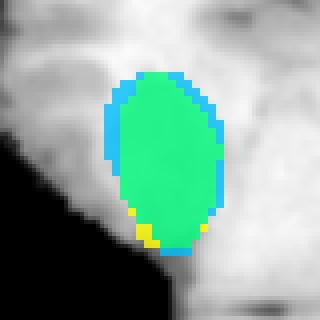}&
    \includegraphics[width=0.13\linewidth]{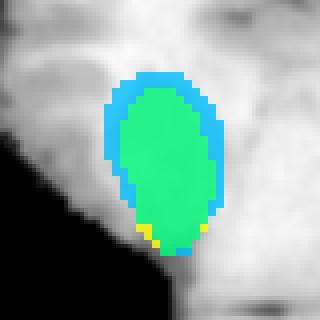}&
    \includegraphics[width=0.13\linewidth]{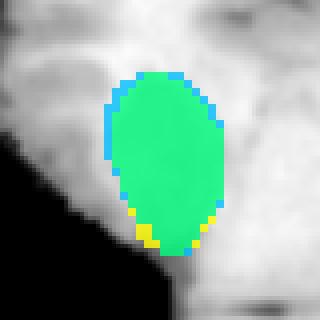}\\
    
    \raisebox{0.1\height}{\rotatebox{90}{\text{Hippocampus}}}&
    \raisebox{0.32\height}{\rotatebox{90}{\text{Surface Z}}}&
    \includegraphics[width=0.13\linewidth]{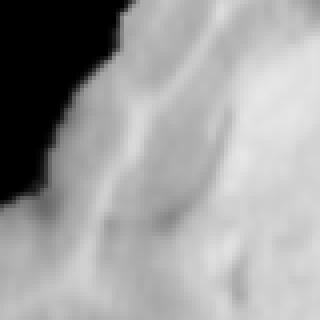}&
    \includegraphics[width=0.13\linewidth]{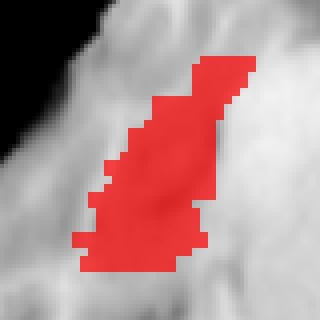}&
    \includegraphics[width=0.13\linewidth]{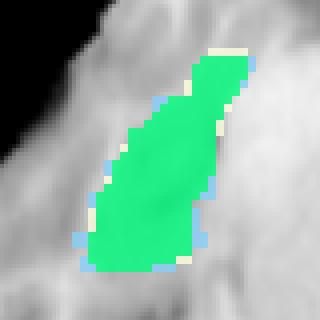}&
    \includegraphics[width=0.13\linewidth]{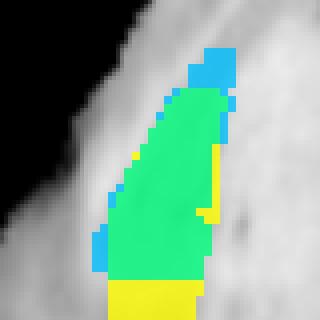}&
    \includegraphics[width=0.13\linewidth]{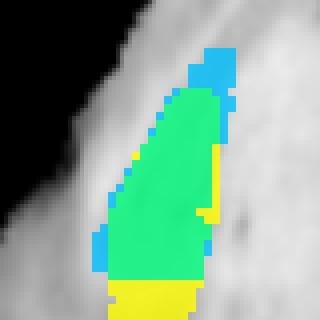}&
    \includegraphics[width=0.13\linewidth]{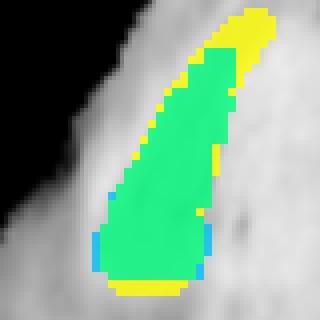}&
    \includegraphics[width=0.13\linewidth]{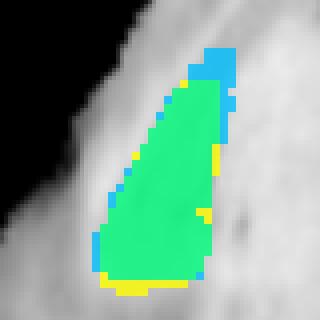}\\

    \end{tabularx}
    \caption{\small{Qualitative results of our framework over three datasets. 
    As shown in the figure, four colors exist in the segmentation results. {\em red} indicates ground-truth. {\em green} indicates correct predictions. {\em blue} indicates predictions where should have been predicted. {\em yellow} indicates wrongly-predicted pixels. 
    Our framework is built on BoxInst~\cite{tian2021boxinst} which is the baseline method. We then compare several methods, including fully-supervised results, the proposed method and demonstrate the high quality of the proposal.}}
    \label{fig:main_res}
\end{figure*}

\textbf{Qualitative Results.} We present the qualitative comparison with two state-of-the-art weakly-supervised segmentation methods in Fig~\ref{fig:main_res}. Compared with BoxInst~\cite{tian2021boxinst}, which is the baseline of our framework, we well preserve the proposal's inner structure detail and outer shape. By comparing Column.\#4 and Column.\#7, the segmentation accuracy is largely improved compared to the baseline in Fig.~\ref{fig:main_res}. This is because the proposed geometric prior and the contrastive similarity help better segment organs. Additionally, the quality of our result is much better than other state-of-the-art methods, especially in terms of internal details. Their methods are based on Multiple Instance Learning (\emph{i.e.}, MIL, GMIL), which mainly focuses on the outer shape of the proposal ignoring the hollows inside the organs. In contrast to their methods, we learn the inner structure and geometric details from the given template organ along with the outer shape that boosts the training of the mask head.

\vspace{-0.1cm}
\subsection{Ablation Study}
In this subsection, we first present the component-wise analysis of our framework and then discuss the effectiveness of the proposed geometric prior and contrastive similarity. 

\subsubsection{Robustness}
In this subsection, we mainly discuss the robustness of the proposed weakly-supervised segmentation framework. Our framework is built on BoxInst~\cite{tian2021boxinst}, which is composed of Backbone and FPN. We analyze the impact of the backbone on the weakly-supervised segmentation performance. As shown in Tab.~\ref{tab:robuseness}, our method is robust to other backbones ({\em i.e.}, ResNet~\cite{he2016deep} and U-Net~\cite{ronneberger2015u}). Furthermore, we verify the generality of the proposed framework by extending to other weakly-supervised segmentation models. Following Chu \emph{et al.}~\cite{chu2021improving}, we use the Ai+L~\cite{chu2021improving} model with a ResNet50~\cite{he2016deep} backbone that is widely used in both non-medical~\cite{lateef2019survey} and medical~\cite{anderson2021automated} segmentation tasks. We concatenate the geometric prior branch and contrastive similarity branch after the decoder. As shown in Tab.~\ref{tab:robuseness}, the baseline model obtains 67.2\% in the KiTS dataset. The segmentation performance is largely improved by 4.9\% after adding the proposed geometric prior and contrastive similarity losses. 

\begin{table}[t!]
\small
\centering
\caption{\small{Robustness analysis. We measure the corresponding dice score on the KiTS21 dataset. We implement the 3D version of BoxInst~\cite{tian2021boxinst} based on its 2D version public code.}}
\renewcommand\arraystretch{1.4}
\newcolumntype{P}[1]{>{\centering\arraybackslash}p{#1}}
\scalebox{0.98}{
	\begin{tabular}{cP{2.1cm}P{2.1cm}cc}
		\specialrule{.8pt}{0pt}{2pt}
		\multirow{2}{*}{\#}& \multirow{2}{*}{Backbone} & \multirow{2}{*}{Model} & Without & With\\
		& & & Our Loss&Our Loss\\
		\specialrule{.4pt}{2pt}{0pt}
		1&ResNet~\cite{he2016deep}&BoxInst 3D~\cite{tian2021boxinst}& 47.8 & \textbf{77.1}\\
		2&UNet~\cite{ronneberger2015u}&BoxInst 3D~\cite{tian2021boxinst}& 49.1& \textbf{80.2}\\
		3&ResNet~\cite{he2016deep}&Ai+L~\cite{chu2021improving}&67.2 & \textbf{72.1}\\
		
		\specialrule{.8pt}{0pt}{2pt}
	\end{tabular}
}
\label{tab:robuseness}
\vspace{-0.6cm}
\end{table}





\subsubsection{Geometric prior}
\begin{figure}[!tb]
    \setlength\tabcolsep{1.6pt}
    \centering
    \small
    \begin{tabularx}{\textwidth}{cccc}
    
    \raisebox{1.2\height}{\rotatebox{90}{\text{Input}}}&
    \includegraphics[width=0.3\linewidth]{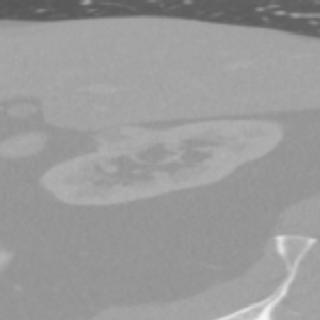}&
    \includegraphics[width=0.3\linewidth]{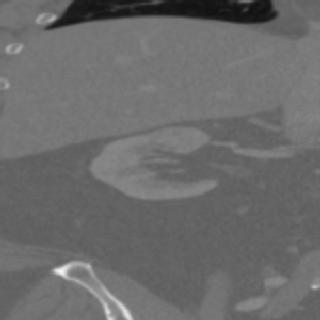}&
    \includegraphics[width=0.3\linewidth]{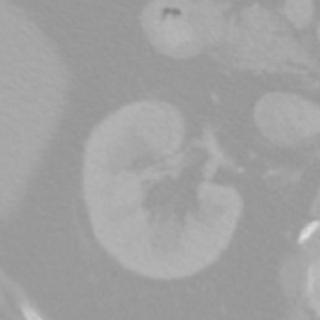}\\
    
    \raisebox{0.8\height}{\rotatebox{90}{\text{volume}}}&
    \includegraphics[width=0.3\linewidth]{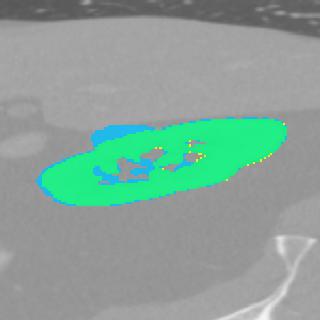}&
    \includegraphics[width=0.3\linewidth]{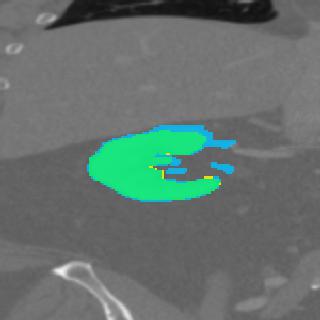}&
    \includegraphics[width=0.3\linewidth]{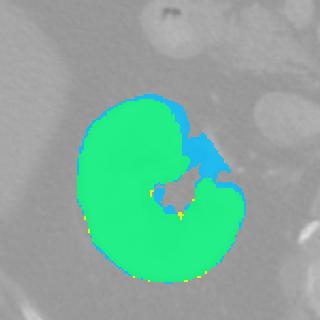}\\
    
    \raisebox{0.35\height}{\rotatebox{90}{\text{point cloud}}}&
    \includegraphics[width=0.3\linewidth]{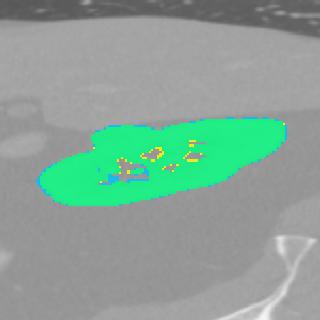}&
    \includegraphics[width=0.3\linewidth]{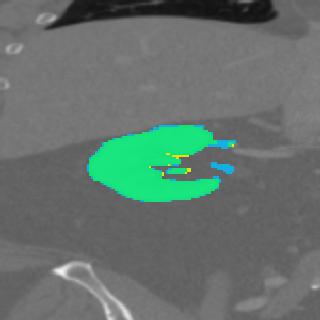}&
    \includegraphics[width=0.3\linewidth]{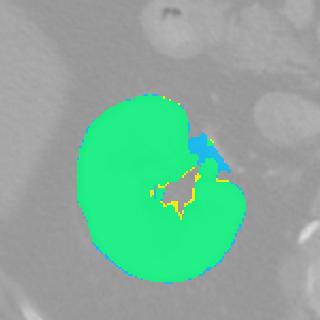}\\

    \end{tabularx}
    \caption{\small{Volume {\em v.s.} Point cloud. We optimize the learning of the geometric shape in the form of volume representation by Dice Loss. As shown in the figure, the proposed point cloud representation largely improves the segmentation result.} }
    \label{fig:represent}
\end{figure}

In the proposed geometric prior, we minimize the Chamfer Distance between the template organ and the proposal in the point cloud embedding space. The point cloud representation is much more flexible than the volume representation. This is because the volume representation is constrained by the uniform voxel grids. The representative point of each grid is fixed to the center of each corresponding grid. In contrast, there is no such constraint in the point cloud representation. The coordinates of points are much more precise and flexible elastic ({\em i.e.}, \{0.17, 0.18, ...\} \emph{v.s.} \{1.00, 2.00, ...\}). This helps the point cloud representation to better describe delicate and complex structures than the conventionally-used volume representation. 
Thus, to validate this point, we conduct an ablation of registration and optimization with the volume representation. More specifically, we first register the organ by the widely-used SimpleITK~\cite{lowekamp2013design}. Here, the registration settings are set as follows: Mean Square metric, random sampling strategy with a percentage of 0.01, shrink factors $[4, 2, 1]$, smoothing sigmas $[2, 1, 0]$, and 100 iterations.
After the registration, we employ the Dice loss to optimize the distance between the segmentation result and the geometric prior. As shown in Fig.~\ref{fig:represent}, the result shows that the point cloud representation can handle more complex architectures and precise geometric shapes, by improving 4.3\% in the Dice Score Coefficient.

\begin{table}[!b]
	\small
	\vspace{-0.6cm}
	\centering
	\caption{{Analysis of geometric prior. $\dagger$ indicates the inner structure of the template.}}
	\renewcommand\arraystretch{1.4}
	\newcolumntype{P}[1]{>{\centering\arraybackslash}p{#1}}
    \resizebox{0.97\linewidth}{!}{
    		\begin{tabular}{cP{4.7cm}P{2.3cm}}
    			\specialrule{.8pt}{0pt}{2pt}
    			\multirow{2}{*}{\#}& \multirow{2}{*}{Ablation Setting} & DSC on \\
		        & &  KiTS \\
    			\specialrule{.4pt}{2pt}{0pt}
    			1 & {w/o Completeness Head} &\multicolumn{1}{c}{79.1}\\
    			2 & w/o $\text{Internal Details}^{\dagger}$ &\multicolumn{1}{c}{76.3} \\
    			\rowcolor[gray]{.9}
    			3 & Baseline & 80.2 \\
    			\specialrule{.8pt}{0pt}{2pt}
    		\end{tabular}
	}
	\label{tab:cpl_internal_geo}
	\normalsize
\end{table}
\newcommand{\setgray}{\cellcolor[gray]{.9}}
\begin{table}[!b]
	\small
	\centering
	\vspace{-0.3cm}
	\caption{{Analysis of the contrastive head. $\dagger$: C stands for the contrastive similarity, S indicates the SSIM similarity and G stands for the grayscale similarity. The inference speed indicates the total time of 1000 runs.}}
	\renewcommand\arraystretch{1.4}
	\newcolumntype{P}[1]{>{\centering\arraybackslash}p{#1}}
    \resizebox{1.\linewidth}{!}{
    		\begin{tabular}{cP{3.5cm}cccc}
    			\specialrule{.8pt}{0pt}{2pt}
    			\multirow{2}{*}{\#} & \multirow{2}{*}{Component Setting}  & Embedding & Embedding & Inference & DSC (\%) \\
    			&  & Dimension &Space$^\dagger$ & Speed & on KiTS\\
                \specialrule{.4pt}{2pt}{0pt}
    			 1 & \multirow{4}{*}{Contrastive Head} &  8 &  C &  3s &  78.4\\
    			2 & &16 & C & 5s & 79.1\\
    			 3 & & 32 &  C &  8s &  80.2\\
    			4 & &64 & C & 15s & 80.4\\
    			\specialrule{.4pt}{2pt}{0pt}
    			 5 & \multirow{3}{*}{Similarity Metrics} & 32&  C &  8s &  80.2\\
    			6 & &32 & G & 8s & 73.4\\
    			 7 & & 32 &  S &  12s &  47.2\\
    			\specialrule{.8pt}{0pt}{2pt}
    		\end{tabular}
	}
	\label{tab:cpl_internal_cons}
	\normalsize
\end{table}

\begin{figure}[!tb]
    \setlength\tabcolsep{1.6pt}
    \centering
    \small
    \begin{tabularx}{\textwidth}{cccc}
    
    \raisebox{1.2\height}{\rotatebox{90}{\text{Input}}}&
    \includegraphics[width=0.3\linewidth]{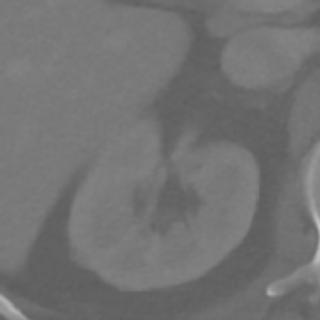}&
    \includegraphics[width=0.3\linewidth]{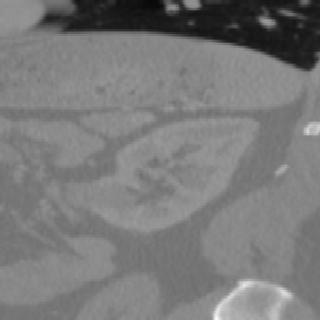}&
    \includegraphics[width=0.3\linewidth]{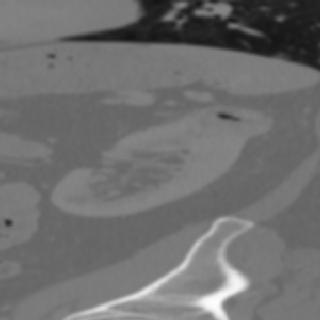}\\
    
    \raisebox{0.38\height}{\rotatebox{90}{\text{w/o internal}}}&
    \includegraphics[width=0.3\linewidth]{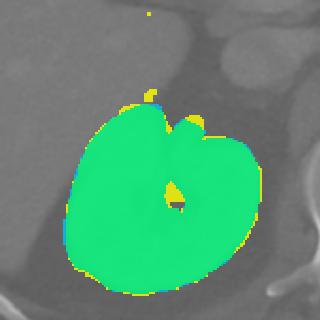}&
    \includegraphics[width=0.3\linewidth]{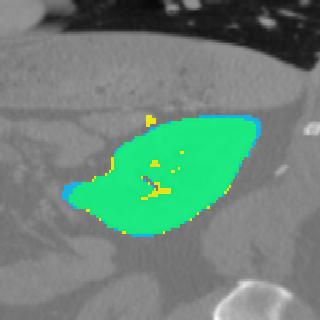}&
    \includegraphics[width=0.3\linewidth]{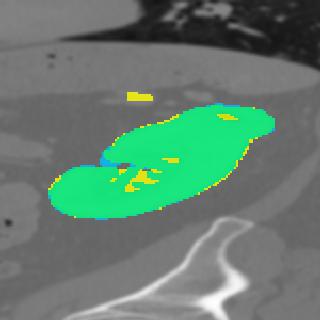}\\
    
    \raisebox{0.38\height}{\rotatebox{90}{\text{w internal}}}&
    \includegraphics[width=0.3\linewidth]{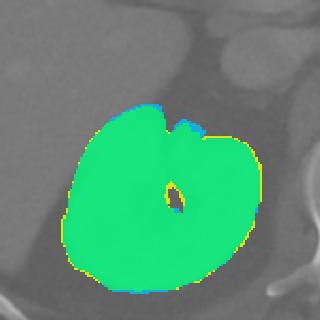}&
    \includegraphics[width=0.3\linewidth]{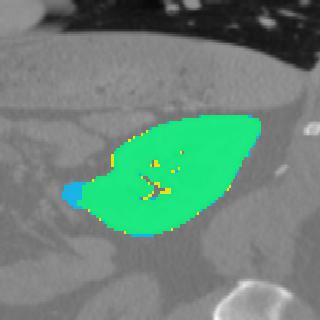}&
    \includegraphics[width=0.3\linewidth]{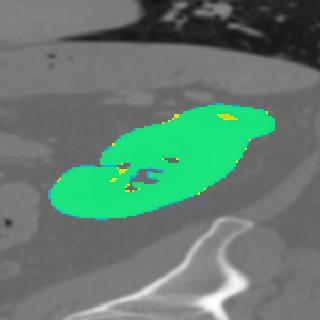}\\

    \end{tabularx}
    \caption{\small{Internal details. To verify the effectiveness of the internal details of the template, we erase the intra-organ variabilities by fulfilling the template organ. As shown in the figure, the segmentation performance drastically degrades especially for organs with delicate and complex structures. Figures from left to right are more and more delicate and complex.} }
    \label{fig:internal}
    \vspace{-0.6cm}
\end{figure}

\textbf{Internal details}
We further analyze the impact of the template from the perspective of outer shape and inner structures, respectively. The experiments in this subsection are organized as follows: We first fulfill the inner structure of the template and only retain the outer shape. Then we train the mask head with the deformed template. Normally, there exists intra-organ variabilities, \emph{i.e.}, as shown in Fig.~\ref{fig:internal}, especially meticulous architectures inside the kidney. And the segmentation performance drastically degrades (3.9\%) for the lack of internal prior. 


\textbf{Analysis of the completeness head}
As discussed in Sec.~\ref{sec:overall}, following nnUNet~\cite{isensee2021nnu}, for an input image, we first partition it into patches and sequentially process these patches. However, during the partitioning, organs may not be complete, which may cause the failure of the learning from the template. Thus, to ensure the success of learning from the template, the proposed completeness head detects whether the proposal in the sampled patch is complete. Only complete proposals are further processed. As shown in Tab.~\ref{tab:cpl_internal_geo}, if we remove the completeness head, the performance degrades by 1.1\% compared with our framework. 

\subsubsection{Architectures of contrastive head}
We conduct experiments on the design of the proposed contrastive head in this subsection. Following previous works~\cite{chen2020simple, hu2021semi}, the contrastive head is composed of a two-layer point-wise convolution that helps distinguish organ tissues from non-organ tissues. To further analyze the structure of the contrastive head, we conduct ablation experiments on the dimension of the embedding space. As shown in Tab.~\ref{tab:cpl_internal_cons}, we evaluate both the performance and the efficiency of different embedding dimension that by considering the trade-off between efficiency and performance, we empirically set the embedding dimension to 32.

\textbf{Contrastive similarity}
 We propose the contrastive similarity in the embedding space. To compare the embedding space with the commonly used gray space, we select the following two similarity metrics: \textit{MSE} and \textit{SSIM}~\cite{wang2004image}. Both \textit{MSE} and \textit{SSIM} are calculated on gray values since medical images are in grayscale. {As shown in Tab.~\ref{tab:cpl_internal_cons}, the \textit{MSE} metric fails to distinguish low-contrast tissues,} and the \textit{SSIM} metric is inferior to the contrastive similarity in terms of both segmentation performance and evaluation efficiency.

\section{Conclusion}
\label{sec:conclusion}
\IEEEPARstart{I}{n} this work, we propose a novel weakly-supervised segmentation framework with bounding-box annotations. We introduce the geometric prior and the contrastive similarity to address the challenges of \qone\ and \qtwo, enhancing the practicability and robustness of the weakly-supervised segmentation framework. The geometric prior enables the learning of delicate and complex structures, and the contrastive similarity helps better distinguish organ pixels from non-organ pixels. 

Our framework is general, which can be easily applied to many weakly-supervised segmentation models and thus improve their performances. It will probably serve as a base for possible future studies on weakly-supervised segmentation, especially for medical image segmentation on organs that are with a specific shape. Extensive experiments are conducted to verify the effectiveness and the superiority of our proposed geometric prior and contrastive similarity.

\bibliographystyle{IEEEtran}
\bibliography{ref}

\begin{thebibliography}{10}
\providecommand{\url}[1]{#1}
\csname url@samestyle\endcsname
\providecommand{\newblock}{\relax}
\providecommand{\bibinfo}[2]{#2}
\providecommand{\BIBentrySTDinterwordspacing}{\spaceskip=0pt\relax}
\providecommand{\BIBentryALTinterwordstretchfactor}{4}
\providecommand{\BIBentryALTinterwordspacing}{\spaceskip=\fontdimen2\font plus
\BIBentryALTinterwordstretchfactor\fontdimen3\font minus
  \fontdimen4\font\relax}
\providecommand{\BIBforeignlanguage}[2]{{%
\expandafter\ifx\csname l@#1\endcsname\relax
\typeout{** WARNING: IEEEtran.bst: No hyphenation pattern has been}%
\typeout{** loaded for the language `#1'. Using the pattern for}%
\typeout{** the default language instead.}%
\else
\language=\csname l@#1\endcsname
\fi
#2}}
\providecommand{\BIBdecl}{\relax}
\BIBdecl

\bibitem{ronneberger2015u}
O.~Ronneberger, P.~Fischer, and T.~Brox, ``U-net: Convolutional networks for
  biomedical image segmentation,'' in \emph{International Conference on Medical
  image computing and computer-assisted intervention}.\hskip 1em plus 0.5em
  minus 0.4em\relax Springer, 2015, pp. 234--241.

\bibitem{isensee2021nnu}
F.~Isensee, P.~F. Jaeger, S.~A. Kohl, J.~Petersen, and K.~H. Maier-Hein,
  ``nnu-net: a self-configuring method for deep learning-based biomedical image
  segmentation,'' \emph{Nature methods}, vol.~18, no.~2, pp. 203--211, 2021.

\bibitem{pathak2015constrained}
D.~Pathak, P.~Krahenbuhl, and T.~Darrell, ``Constrained convolutional neural
  networks for weakly supervised segmentation,'' in \emph{Proceedings of the
  IEEE international conference on computer vision}, 2015, pp. 1796--1804.

\bibitem{jia2017constrained}
Z.~Jia, X.~Huang, I.~Eric, C.~Chang, and Y.~Xu, ``Constrained deep weak
  supervision for histopathology image segmentation,'' \emph{IEEE transactions
  on medical imaging}, vol.~36, no.~11, pp. 2376--2388, 2017.

\bibitem{kervadec2019constrained}
H.~Kervadec, J.~Dolz, M.~Tang, E.~Granger, Y.~Boykov, and I.~B. Ayed,
  ``Constrained-cnn losses for weakly supervised segmentation,'' \emph{Medical
  image analysis}, vol.~54, pp. 88--99, 2019.

\bibitem{bateson2019constrained}
M.~Bateson, H.~Kervadec, J.~Dolz, H.~Lombaert, and I.~B. Ayed, ``Constrained
  domain adaptation for segmentation,'' in \emph{International Conference on
  Medical Image Computing and Computer-Assisted Intervention}.\hskip 1em plus
  0.5em minus 0.4em\relax Springer, 2019, pp. 326--334.

\bibitem{tian2021boxinst}
Z.~Tian, C.~Shen, X.~Wang, and H.~Chen, ``Boxinst: High-performance instance
  segmentation with box annotations,'' in \emph{Proceedings of the IEEE/CVF
  Conference on Computer Vision and Pattern Recognition}, 2021, pp. 5443--5452.

\bibitem{budrys2018artifacts}
T.~Budrys, V.~Veikutis, S.~Lukosevicius, R.~Gleizniene, E.~Monastyreckiene, and
  I.~Kulakiene, ``Artifacts in magnetic resonance imaging: how it can really
  affect diagnostic image quality and confuse clinical diagnosis?''
  \emph{Journal of Vibroengineering}, vol.~20, no.~2, pp. 1202--1213, 2018.

\bibitem{boas2012ct}
F.~E. Boas, D.~Fleischmann \emph{et~al.}, ``Ct artifacts: causes and reduction
  techniques,'' \emph{Imaging Med}, vol.~4, no.~2, pp. 229--240, 2012.

\bibitem{sarkar2018subjective}
S.~N. Sarkar, D.~B. Hackney, R.~L. Greenman, B.~A. Vachha, E.~A. Johnson,
  S.~Nagle, and G.~Moonis, ``A subjective and objective comparison of tissue
  contrast and imaging artifacts present in routine spin echoes and in
  iterative decomposition of asymmetric spin echoes for soft tissue neck mri,''
  \emph{European journal of radiology}, vol. 102, pp. 202--207, 2018.

\bibitem{xie2020grnet}
H.~Xie, H.~Yao, S.~Zhou, J.~Mao, S.~Zhang, and W.~Sun, ``Grnet: Gridding
  residual network for dense point cloud completion,'' in \emph{European
  Conference on Computer Vision}.\hskip 1em plus 0.5em minus 0.4em\relax
  Springer, 2020, pp. 365--381.

\bibitem{chaitanya2020contrastive}
K.~Chaitanya, E.~Erdil, N.~Karani, and E.~Konukoglu, ``Contrastive learning of
  global and local features for medical image segmentation with limited
  annotations,'' \emph{Advances in Neural Information Processing Systems},
  vol.~33, pp. 12\,546--12\,558, 2020.

\bibitem{chu2021improving}
T.~Chu, X.~Li, H.~V. Vo, R.~M. Summers, and E.~Sizikova, ``Improving weakly
  supervised lesion segmentation using multi-task learning,'' in \emph{Medical
  Imaging with Deep Learning}.\hskip 1em plus 0.5em minus 0.4em\relax PMLR,
  2021, pp. 60--73.

\bibitem{lin2016scribblesup}
D.~Lin, J.~Dai, J.~Jia, K.~He, and J.~Sun, ``Scribblesup: Scribble-supervised
  convolutional networks for semantic segmentation,'' in \emph{Proceedings of
  the IEEE conference on computer vision and pattern recognition}, 2016, pp.
  3159--3167.

\bibitem{bearman2016s}
A.~Bearman, O.~Russakovsky, V.~Ferrari, and L.~Fei-Fei, ``What’s the point:
  Semantic segmentation with point supervision,'' in \emph{European conference
  on computer vision}.\hskip 1em plus 0.5em minus 0.4em\relax Springer, 2016,
  pp. 549--565.

\bibitem{Qu2019}
H.~{Qu} \emph{et~al.}, ``{Weakly Supervised Deep Nuclei Segmentation Using
  Partial Points Annotation in Histopathology Images},'' \emph{IEEE
  Transactions on Medical Imaging}, pp. 1--1, 2020.

\bibitem{patel2022weakly}
G.~Patel and J.~Dolz, ``Weakly supervised segmentation with cross-modality
  equivariant constraints.'' \emph{Medical Image Analysis}, p. 102374, 2022.

\bibitem{Xu}
G.~{Xu} \emph{et~al.}, ``{CAMEL: A Weakly Supervised Learning Framework for
  Histopathology Image Segmentation},'' in \emph{{2019 IEEE/CVF International
  Conference on Computer Vision (ICCV)}}, 2019, pp. 10\,681--10\,690.

\bibitem{wang2019weakly2}
X.~Wang \emph{et~al.}, ``{Weakly Supervised Deep Learning for Whole Slide Lung
  Cancer Image Analysis},'' \emph{IEEE Transactions on Cybernetics}, 2019.

\bibitem{Grady2006Random}
L.~Grady, ``{Random Walks for Image Segmentation},'' \emph{IEEE Transactions on
  Pattern Analysis and Machine Intelligence}, vol.~28, pp. 1768--1783, 2006.

\bibitem{dai2015boxsup}
J.~Dai, K.~He, and J.~Sun, ``Boxsup: Exploiting bounding boxes to supervise
  convolutional networks for semantic segmentation,'' in \emph{Proceedings of
  the IEEE international conference on computer vision}, 2015, pp. 1635--1643.

\bibitem{papandreou2015weakly}
G.~Papandreou, L.-C. Chen, K.~P. Murphy, and A.~L. Yuille, ``Weakly-and
  semi-supervised learning of a deep convolutional network for semantic image
  segmentation,'' in \emph{Proceedings of the IEEE international conference on
  computer vision}, 2015, pp. 1742--1750.

\bibitem{khoreva2017simple}
A.~Khoreva, R.~Benenson, J.~Hosang, M.~Hein, and B.~Schiele, ``Simple does it:
  Weakly supervised instance and semantic segmentation,'' in \emph{Proceedings
  of the IEEE conference on computer vision and pattern recognition}, 2017, pp.
  876--885.

\bibitem{pu2018graphnet}
M.~Pu, Y.~Huang, Q.~Guan, and Q.~Zou, ``Graphnet: Learning image pseudo
  annotations for weakly-supervised semantic segmentation,'' in
  \emph{Proceedings of the 26th ACM international conference on Multimedia},
  2018, pp. 483--491.

\bibitem{rajchl2016deepcut}
M.~Rajchl, M.~C. Lee, O.~Oktay, K.~Kamnitsas, J.~Passerat-Palmbach, W.~Bai,
  M.~Damodaram, M.~A. Rutherford, J.~V. Hajnal, B.~Kainz \emph{et~al.},
  ``Deepcut: Object segmentation from bounding box annotations using
  convolutional neural networks,'' \emph{IEEE transactions on medical imaging},
  vol.~36, no.~2, pp. 674--683, 2016.

\bibitem{patenaude2011bayesian}
B.~Patenaude, S.~M. Smith, D.~N. Kennedy, and M.~Jenkinson, ``A bayesian model
  of shape and appearance for subcortical brain segmentation,''
  \emph{Neuroimage}, vol.~56, no.~3, pp. 907--922, 2011.

\bibitem{sabuncu2010generative}
M.~R. Sabuncu, B.~T. Yeo, K.~Van~Leemput, B.~Fischl, and P.~Golland, ``A
  generative model for image segmentation based on label fusion,'' \emph{IEEE
  transactions on medical imaging}, vol.~29, no.~10, pp. 1714--1729, 2010.

\bibitem{fischl2002whole}
B.~Fischl, D.~H. Salat, E.~Busa, M.~Albert, M.~Dieterich, C.~Haselgrove, A.~Van
  Der~Kouwe, R.~Killiany, D.~Kennedy, S.~Klaveness \emph{et~al.}, ``Whole brain
  segmentation: automated labeling of neuroanatomical structures in the human
  brain,'' \emph{Neuron}, vol.~33, no.~3, pp. 341--355, 2002.

\bibitem{iglesias2015multi}
J.~E. Iglesias and M.~R. Sabuncu, ``Multi-atlas segmentation of biomedical
  images: a survey,'' \emph{Medical image analysis}, vol.~24, no.~1, pp.
  205--219, 2015.

\bibitem{gao2016segmentation}
M.~Gao, Z.~Xu, L.~Lu, A.~Wu, I.~Nogues, R.~M. Summers, and D.~J. Mollura,
  ``Segmentation label propagation using deep convolutional neural networks and
  dense conditional random field,'' in \emph{2016 IEEE 13th International
  Symposium on Biomedical Imaging (ISBI)}.\hskip 1em plus 0.5em minus
  0.4em\relax IEEE, 2016, pp. 1265--1268.

\bibitem{ganaye2018semi}
P.-A. Ganaye, M.~Sdika, and H.~Benoit-Cattin, ``Semi-supervised learning for
  segmentation under semantic constraint,'' in \emph{International Conference
  on Medical Image Computing and Computer-Assisted Intervention}.\hskip 1em
  plus 0.5em minus 0.4em\relax Springer, 2018, pp. 595--602.

\bibitem{bentaieb2016topology}
A.~BenTaieb and G.~Hamarneh, ``Topology aware fully convolutional networks for
  histology gland segmentation,'' in \emph{International conference on medical
  image computing and computer-assisted intervention}.\hskip 1em plus 0.5em
  minus 0.4em\relax Springer, 2016, pp. 460--468.

\bibitem{chen2017dcan}
H.~Chen, X.~Qi, L.~Yu, Q.~Dou, J.~Qin, and P.-A. Heng, ``Dcan: Deep
  contour-aware networks for object instance segmentation from histology
  images,'' \emph{Medical image analysis}, vol.~36, pp. 135--146, 2017.

\bibitem{zhou2019prior}
Y.~Zhou, Z.~Li, S.~Bai, C.~Wang, X.~Chen, M.~Han, E.~Fishman, and A.~L. Yuille,
  ``Prior-aware neural network for partially-supervised multi-organ
  segmentation,'' in \emph{Proceedings of the IEEE/CVF International Conference
  on Computer Vision}, 2019, pp. 10\,672--10\,681.

\bibitem{oktay2017anatomically}
O.~Oktay, E.~Ferrante, K.~Kamnitsas, M.~Heinrich, W.~Bai, J.~Caballero, S.~A.
  Cook, A.~De~Marvao, T.~Dawes, D.~P. O‘Regan \emph{et~al.}, ``Anatomically
  constrained neural networks (acnns): application to cardiac image enhancement
  and segmentation,'' \emph{IEEE transactions on medical imaging}, vol.~37,
  no.~2, pp. 384--395, 2017.

\bibitem{dalca2018anatomical}
A.~V. Dalca, J.~Guttag, and M.~R. Sabuncu, ``Anatomical priors in convolutional
  networks for unsupervised biomedical segmentation,'' in \emph{Proceedings of
  the IEEE Conference on Computer Vision and Pattern Recognition}, 2018, pp.
  9290--9299.

\bibitem{xie2020pointcontrast}
S.~Xie, J.~Gu, D.~Guo, C.~R. Qi, L.~Guibas, and O.~Litany, ``Pointcontrast:
  Unsupervised pre-training for 3d point cloud understanding,'' in
  \emph{European conference on computer vision}.\hskip 1em plus 0.5em minus
  0.4em\relax Springer, 2020, pp. 574--591.

\bibitem{xie2021propagate}
Z.~Xie, Y.~Lin, Z.~Zhang, Y.~Cao, S.~Lin, and H.~Hu, ``Propagate yourself:
  Exploring pixel-level consistency for unsupervised visual representation
  learning,'' in \emph{Proceedings of the IEEE/CVF Conference on Computer
  Vision and Pattern Recognition}, 2021, pp. 16\,684--16\,693.

\bibitem{he2020momentum}
K.~He, H.~Fan, Y.~Wu, S.~Xie, and R.~Girshick, ``Momentum contrast for
  unsupervised visual representation learning,'' in \emph{Proceedings of the
  IEEE/CVF conference on computer vision and pattern recognition}, 2020, pp.
  9729--9738.

\bibitem{van2021unsupervised}
W.~Van~Gansbeke, S.~Vandenhende, S.~Georgoulis, and L.~Van~Gool, ``Unsupervised
  semantic segmentation by contrasting object mask proposals,'' in
  \emph{Proceedings of the IEEE/CVF International Conference on Computer
  Vision}, 2021, pp. 10\,052--10\,062.

\bibitem{wang2021exploring}
W.~Wang, T.~Zhou, F.~Yu, J.~Dai, E.~Konukoglu, and L.~Van~Gool, ``Exploring
  cross-image pixel contrast for semantic segmentation,'' in \emph{Proceedings
  of the IEEE/CVF International Conference on Computer Vision}, 2021, pp.
  7303--7313.

\bibitem{hu2021semi}
X.~Hu, D.~Zeng, X.~Xu, and Y.~Shi, ``Semi-supervised contrastive learning for
  label-efficient medical image segmentation,'' in \emph{International
  Conference on Medical Image Computing and Computer-Assisted
  Intervention}.\hskip 1em plus 0.5em minus 0.4em\relax Springer, 2021, pp.
  481--490.

\bibitem{Zhou2018}
Q.-Y. Zhou, J.~Park, and V.~Koltun, ``{Open3D}: {A} modern library for {3D}
  data processing,'' \emph{arXiv:1801.09847}, 2018.

\bibitem{jang2016categorical}
E.~Jang, S.~Gu, and B.~Poole, ``Categorical reparameterization with
  gumbel-softmax,'' \emph{arXiv preprint arXiv:1611.01144}, 2016.

\bibitem{chen2020simple}
T.~Chen, S.~Kornblith, M.~Norouzi, and G.~Hinton, ``A simple framework for
  contrastive learning of visual representations,'' in \emph{International
  conference on machine learning}.\hskip 1em plus 0.5em minus 0.4em\relax PMLR,
  2020, pp. 1597--1607.

\bibitem{simonyan2014very}
K.~Simonyan and A.~Zisserman, ``Very deep convolutional networks for
  large-scale image recognition,'' \emph{arXiv preprint arXiv:1409.1556}, 2014.

\bibitem{Dolz2020}
H.~Kervadec, J.~Dolz, S.~Wang, E.~Granger, and I.~B. Ayed, ``{Bounding Boxes
  for Weakly Supervised Segmentation: Global Constraints Get Close to Full
  Supervision},'' \emph{arXiv Preprint ArXiv:2004.06816}, 2020.

\bibitem{paszke2016enet}
A.~Paszke, A.~Chaurasia, S.~Kim, and E.~Culurciello, ``Enet: A deep neural
  network architecture for real-time semantic segmentation,'' \emph{arXiv
  preprint arXiv:1606.02147}, 2016.

\bibitem{bilic2019liver}
P.~Bilic, P.~F. Christ, E.~Vorontsov, G.~Chlebus, H.~Chen, Q.~Dou, C.-W. Fu,
  X.~Han, P.-A. Heng, J.~Hesser \emph{et~al.}, ``The liver tumor segmentation
  benchmark (lits),'' \emph{arXiv preprint arXiv:1901.04056}, 2019.

\bibitem{heller2020state}
N.~Heller, F.~Isensee, K.~H. Maier-Hein, X.~Hou, C.~Xie, F.~Li, Y.~Nan, G.~Mu,
  Z.~Lin, M.~Han \emph{et~al.}, ``The state of the art in kidney and kidney
  tumor segmentation in contrast-enhanced ct imaging: Results of the kits19
  challenge,'' \emph{Medical Image Analysis}, p. 101821, 2020.

\bibitem{he2016deep}
K.~He, X.~Zhang, S.~Ren, and J.~Sun, ``Deep residual learning for image
  recognition,'' in \emph{Proceedings of the IEEE conference on computer vision
  and pattern recognition}, 2016, pp. 770--778.

\bibitem{lateef2019survey}
F.~Lateef and Y.~Ruichek, ``Survey on semantic segmentation using deep learning
  techniques,'' \emph{Neurocomputing}, vol. 338, pp. 321--348, 2019.

\bibitem{anderson2021automated}
B.~M. Anderson, E.~Y. Lin, C.~E. Cardenas, D.~A. Gress, W.~D. Erwin, B.~C.
  Odisio, E.~J. Koay, and K.~K. Brock, ``Automated contouring of contrast and
  noncontrast computed tomography liver images with fully convolutional
  networks,'' \emph{Advances in radiation oncology}, vol.~6, no.~1, p. 100464,
  2021.

\bibitem{lowekamp2013design}
B.~C. Lowekamp, D.~T. Chen, L.~Ib{\'a}{\~n}ez, and D.~Blezek, ``The design of
  simpleitk,'' \emph{Frontiers in neuroinformatics}, vol.~7, p.~45, 2013.

\bibitem{wang2004image}
Z.~Wang, A.~C. Bovik, H.~R. Sheikh, and E.~P. Simoncelli, ``Image quality
  assessment: from error visibility to structural similarity,'' \emph{IEEE
  transactions on image processing}, vol.~13, no.~4, pp. 600--612, 2004.

\end{thebibliography}

\end{document}